\documentclass[conference]{IEEEtran}

\IEEEoverridecommandlockouts
\usepackage{cite}
\usepackage{amsmath,amssymb,amsfonts}
\usepackage{graphicx}
\usepackage{textcomp}
\usepackage{xcolor}
\usepackage{graphicx}
\usepackage[show]{chato-notes}
\usepackage{amsmath,amssymb}
\usepackage{epsfig}
\usepackage{algorithm}

\usepackage[noend]{algorithmic}

\usepackage{url}
\usepackage{hyperref}

\hypersetup{
    bookmarks=true,         
    unicode=false,          
    pdftoolbar=true,        
    pdfmenubar=true,        
    pdffitwindow=false,     
    pdfstartview={FitH},    
    pdftitle={My title},    
    pdfauthor={Author},     
    pdfsubject={Subject},   
    pdfcreator={Creator},   
    pdfproducer={Producer}, 
    pdfnewwindow=true,      
    colorlinks=true,       
    linkcolor=black,          
    citecolor=black,        
    filecolor=black,      
    urlcolor=black           
}

\newcommand{\mindiam}{\textsc{CSO-MinDiam-MinDeg}} 
\newcommand{\mindegdist}{\textsc{CSO-MaxMinDeg-Dist}} 
\newcommand{\mindeg}{\textsc{CSO-MaxMinDeg-Diam}} 

\newcommand{\amindegdist}{\textsc{Algo-MaxMinDeg-Dist}}
\newcommand{\amindiam}{\textsc{Algo-MinDiam-MinDeg}}
\newcommand{\amindeg}{\textsc{Algo-MaxMinDeg-Diam}}

\newcommand{\spara}[1]{\smallskip\noindent{\bf #1}}

\newtheorem{problem}{Problem}

\newcommand{\NP}{$\mathbf{NP}$}
\newcommand{\NPhard}{$\mathbf{NP}$-hard}

\newcommand{\squishlist}{
 \begin{list}{$\bullet$}
  {  \setlength{\itemsep}{0pt}
     \setlength{\parsep}{3pt}
     \setlength{\topsep}{3pt}
     \setlength{\partopsep}{0pt}
     \setlength{\leftmargin}{2em}
     \setlength{\labelwidth}{1.5em}
     \setlength{\labelsep}{0.5em}
} }
\newcommand{\squishlisttight}{
 \begin{list}{$\bullet$}
  { \setlength{\itemsep}{0pt}
    \setlength{\parsep}{0pt}
    \setlength{\topsep}{0pt}
    \setlength{\partopsep}{0pt}
    \setlength{\leftmargin}{2em}
    \setlength{\labelwidth}{1.5em}
    \setlength{\labelsep}{0.5em}
} }

\newcommand{\squishdesc}{
 \begin{list}{}
  {  \setlength{\itemsep}{0pt}
     \setlength{\parsep}{3pt}
     \setlength{\topsep}{3pt}
     \setlength{\partopsep}{0pt}
     \setlength{\leftmargin}{1em}
     \setlength{\labelwidth}{1.5em}
     \setlength{\labelsep}{0.5em}
} }

\newcommand{\squishend}{
  \end{list}
}

\hypersetup{
	pdftitle={Better Fewer but Better: Community Search with Outliers},
}
\newtheorem{theorem}{Theorem}
\newtheorem{lemma}[theorem]{Lemma}

\newtheorem{corollary}{Corollary}

\def\BibTeX{{\rm B\kern-.05em{\sc i\kern-.025em b}\kern-.08em
    T\kern-.1667em\lower.7ex\hbox{E}\kern-.125emX}}
\begin{document}

\title{Better Fewer but Better:\\ Community Search with Outliers
}

\author{\IEEEauthorblockN{Francesco Bonchi}
\IEEEauthorblockA{\textit{ISI Foundation}, Turin, Italy  \\
\textit{Eurecat}, Barcelona, Spain  \\
francesco.bonchi@isi.it}
\and
\IEEEauthorblockN{Lorenzo Severini}
\IEEEauthorblockA{\textit{UniCredit Services} \\
Rome, Italy \\
loseverini@gmail.com}
\and
\IEEEauthorblockN{Mauro Sozio}
\IEEEauthorblockA{\textit{Telecom Paris, IP Paris }\\
Paris, France \\
sozio@telecom-paris.fr}
}

\maketitle

\begin{abstract}
	Given a set of vertices in a network, that we believe are of interest for the application under analysis, \emph{community search} is the problem of producing a subgraph potentially explaining the relationships existing among the vertices of interest. In practice this means that the solution should add some vertices to the query ones, so to create a connected subgraph that exhibits some ``cohesiveness'' property.
	This problem has received increasing attention in recent years: while several cohesiveness functions have been studied, the bulk of the literature looks for a solution subgraphs containing \emph{all} the query vertices. However, in many exploratory analyses we might only have a reasonable belief about the vertices of interest: if only one of them is not really related to the others, forcing the solution to include all of them might hide the existence of much more cohesive and meaningful subgraphs, that we could have found by allowing the solution to detect and drop the outlier vertex.
	In this paper we study the problem of \emph{community search with outliers}, where we are allowed to drop up to $k$ query vertices, with $k$ being an input parameter. We consider three of the most used measures of cohesiveness: the minimum degree, the diameter of the subgraph and the maximum distance with a query vertex. By optimizing one and using one of the others as a constraint we obtain three optimization problems: we study their hardness and we propose different exact and approximation algorithms.
	
	%
	
\end{abstract}


\section{Introduction}
\label{sec:intro}
\vspace{-1mm}

Community search is a fundamental graph mining problem which has recently received a great deal of attention:
given a graph and a set of query vertices, we wish to find a subset of vertices containing the query ones such that the induced subgraph is connected and optimizes some cohesiveness measure. The extracted subgraph may provide useful insights on the relationships existing among the query vertices and other vertices in the graph. For instance, given a set of suspected terrorists in a social networks, which other individuals in the social network should we monitor? Given a set of proteins of interest in a protein-protein interaction network, which other proteins can participate in pathways with them?
This is an exploratory data analysis task: in most applications we might be given some query vertices that \emph{we believe} might have some interesting connection or participate in some relevant pattern or substructure. However, it might be the case that \emph{not all of them} are relevant for the discovery task at hand. By forcing the solution to connect them all, we might produce much larger and less cohesive solutions, hiding the really interesting ones that we could instead find by just allowing some query vertices not to belong to the solution. While several cohesiveness functions have been studied, the bulk of the literature (briefly surveyed next) enforce that \emph{all} query vertices be present in the output solution. Before presenting our contributions, we provide some background definition and present the relevant related literature.
\label{sec:related}

\spara{Background and related work.} Given a graph $G=(V,E)$ and a set of query vertices $Q \subseteq V$, a wide family of problems requires to find a connected subgraph $H$ of $G$, that contains all query vertices $Q$, while exhibiting some nice properties of cohesiveness, compactness or density. Several problems fit under this framework, such as \emph{community search}~\cite{SozioKDD10,SozioLocalSIGMOD14,BarbieriBGG15},
\emph{seed set expansion} \cite{Kloumann}, and \emph{connectivity subgraphs} \cite{connect,CenterpieceKDD06,ruchansky2015minimum,akoglu2013mining}.
Kloumann and Kleinberg \cite{Kloumann} provide a systematic evaluation of different methods for \emph{seed set expansion} on graphs with known community structure, assuming that the seed set $Q$ is made of vertices belonging to the same community.
Faloutsos et al.~\cite{connect} address the problem of finding a subgraph that
connects two query vertices ($|Q| = 2$) and contains at most $b$ other vertices, optimizing a measure of proximity based on \emph{electrical-current flows}.
Tong and Faloutsos \cite{CenterpieceKDD06} extend \cite{connect} by introducing the concept of \emph{Center-piece Subgraph} dealing with query sets of any size.
Sozio and Gionis~\cite{SozioKDD10} developed a framework for solving a wide range of community search optimization problems. The most basic problem studied in~\cite{SozioKDD10} consists of finding a connected subgraph containing $Q$ and maximizing the minimum degree. They developed an efficient algorithm for solving such a problem optimally,
 however, it is well-known that such a variant of the problem suffers from the \emph{free rider} effect. To alleviate such a problem other constraints on the output graph can be enforced (so-called \emph{monotone} functions). In our work, we focus on the variant where a distance constraint between the nodes in the graph and the query nodes is enforced, additionally. We adapt such a variant to the case with outliers while maintaining the nice property that it can be solved optimally.
More recently, Cui~\emph{et~al.}~\cite{SozioLocalSIGMOD14} devise a local-search approach to improve the efficiency of the method defined in~\cite{SozioKDD10}, but only for the special case of a single query vertex.
The case of multiple query vertices has instead been addressed by Barbieri~\emph{et~al.}~\cite{BarbieriBGG15}, who exploit core decomposition as a preprocessing step to improve efficiency.
Ruchansky et al. \cite{ruchansky2015minimum} introduce the parameter-free problem of extracting the \emph{Minimum Wiener Connector}, that is the connected subgraph containing $Q$ which minimizes the pairwise sum of shortest-path distances among its vertices.
Recent approaches also introduce the flexibility of having query vertices belonging to different communities~\cite{Bian2018,yan2019constrained}.
Finally, community search has been formalized for attributed graphs~\cite{huang2017attribute,fang2017attributed}, spatial graphs~\cite{fang2017spatial} and temporal graphs~\cite{bonchitemp1,bonchitemp2}.
All these approaches look for a connected subgraph of the input graph containing \emph{all} the query vertices. Three recent approaches allow some forms of outliers in community search.
Akoglu et al. \cite{akoglu2013mining} study the problem of finding \emph{pathways}, i.e., connection subgraphs for a large query set $Q$, in terms of the Minimum Description Length (MDL) principle.  According to MDL, a pathway is simple when only a few bits are needed to relay which edges should be followed to visit all of $Q$.
Given a graph $G$ and a query set $Q$, Gionis et al. \cite{gionisbump} study the problem of finding a connected subgraph of $G$ that has more vertices that belong to $Q$ than vertices that do not.
Ruchansky et al. \cite{RuchanskyBGGK17} introduce the problem of finding the \emph{minimum inefficiency subgraph}: they show that the problem is  \NPhard\ and develop an efficient greedy algorithm. The minimum inefficiency subgraph is not required to be connected: as such one could consider the query vertices that ends up disconnected as outliers.
None of these three approaches enforces an upper bound on the number of outliers that be dropped, while no theoretical guarantee is provided. Instead our work studies the general problem with the input parameter on the maximum number of allowed outliers, and presents algorithms with strong theoretical guarantees. 

\spara{Problems studied and results.}
In this paper, we study the \emph{community search with outliers} problem where we are allowed to drop up to $k$ query vertices from the input graph, with $k$ being provided in input. We focus on the most widely-used cohesiveness functions considered in the literature, such as the \emph{diameter} of the output graph~\cite{DBLP:journals/pvldb/HuangLYC15}, and its \emph{minimum degree}~\cite{SozioKDD10}, while studying their generalization to the case with outliers.  By optimizing one measure and using the other as a constraint we obtain two optimization problems. Our work is the first one to propose algorithms with strong theoretical guarantees for the problem of community search with outliers. More in details:

\begin{itemize}

\item For the problem of minimizing the diameter under a minimum degree constraint when at most $k$ query vertex can be dropped, we develop a $2$-approximation algorithm. When $k$ is a constant and there is no constraint on the minimum degree our algorithm boasts almost linear time in the size of the input graph (modulo a logarithmic factor). We then show that there is no algorithm with an approximation guarantee better than $2$, unless $\mathbf{P}$=\NP, making our result tight.

\smallskip 

\item For the problem of maximizing the minimum degree under a diameter constraint when at most $k$ query vertex can be dropped, we show that it \NPhard~and we develop a bicriteria algorithm with approximation guarantees.

\smallskip 

\item Finally, we also study a variant where, instead of a constraint on the diameter (i.e., maximum distance among any pair of vertices in the solution, regardless they are query vertices or not), we have a constraint on the maximum distance between a query vertex and any other vertex in the solution subgraph $H$. We will show that a solution for such a problem can be found efficiently in polynomial time.
\end{itemize}


Our theoretical results are complemented with an extensive experimental evaluation on real-world graphs, confirming the fact that by allowing outliers to be removed, much more cohesive solutions can be found. We  also assess the running time of our algorithms on large real-world graphs while we show their effectiveness in identifying outlier query vertices.



\section{Problem Definitions and Complexity}
\label{sec:problem}
\vspace{-1mm}
We are given an undirected graph $G=(V_G,E_G)$, a set of query vertices $Q \subseteq V_G$, and a positive integer $k < |Q|$.
The shortest-path distance $d_G(u,v)$ between two vertices $u,v$ in $G$ is defined as the length of the shortest path connecting $u,v$ in $G$. In the case when $u,v$ are not in a same connected component we let $d(u,v):= + \infty$. The diameter of a graph $G$, denoted with $diam(G)$, is defined as the maximum shortest-path distance between any two vertices in $G$.
In this section, we shall define the main problems studied in the rest of the paper.  We call our problems \mindeg, \mindegdist, \mindiam, where \textsc{CSO} indicates the general problem (community search with outliers), the first part after '-' indicates the objective function (e.g. \textsc{MinDiam}), while the last part indicates the main constraint (e.g. \textsc{MinDeg}).
We first consider the variant where we wish to minimize the diameter while satisfying a constraint on the minimum degree.  The problem of minimizing the diameter has been studied in~\cite{DBLP:journals/pvldb/HuangLYC15} (without outliers). In our work, we use  a similar proof strategy, while following the framework developed in~\cite{SozioKDD10}.

\smallskip

\begin{problem}[\mindiam]\label{prob:min_dia}
	Given a graph $G=(V_G,E_G)$, a set of query nodes $Q \subseteq V_G$, an integer $k \geq 0$  such that  $k \leq |Q| -1$, an integer $\delta_{\min} \geq 0$, find an induced subgraph $H=(V_H,E_H)$ of $G$  such that:
	
	\begin{enumerate}
		\item $H$ is connected;
		\item $|V_H \cap Q| \geq |Q| - k$;
		\item the minimum degree in $H$ is at least $\delta_{\min}$;
		\item $diam(H)$ is minimized among all subgraphs satisfying 1-3;
	\end{enumerate}
\end{problem}

\smallskip

\begin{theorem}\label{theo:hard1}
	\mindiam\ (Problem \ref{prob:min_dia}) is \NPhard\ even when $\delta_{min}=1$.
\end{theorem}
\begin{IEEEproof}
	We reduce the decision version of Maximum Clique problem (MDC) to \mindiam. Notice that a graph is a clique if and only if its diameter is one. Given a graph $G=(V_G,E_G)$ and an integer $h$, we wish to determine whether $G$ contains a clique of size at least $h$. We construct an instance of MDC problem with $Q=V_G$ and $h=|Q|-k$.
	Let $H=(V_H,E_H)$ be the solution of MDC in this instance. By definition of MDC, $V_H\subseteq V_G$  and $|V_H|\geq |Q|-k$.
	If $diam(H)=1$, since $H$ is and induced subgraph of $G$, $G$ contains a clique of size $h=|Q|-k$.
	On the other hand, consider the case when $diam(H)>1$. Since, by definition, $H$ is the subgraph with size $|Q|-k$ with minimum diameter, this means that $G$ does not contain any clique of size at least $h=|Q|-k$.
\end{IEEEproof}

From the fact that the diameter is an integer, it follows that any approximation algorithm with an approximation guarantee strictly better than $2$ would compute an optimum solution to the MDC. The following corollary follows.

\smallskip

\begin{corollary}
	\mindiam\  cannot be approximated in polynomial time within a factor of $(2-\epsilon)$ for any $\epsilon>0$, unless $\mathbf{P}$=\NP.
\end{corollary}

\smallskip

We next consider the community search problem where the objective is to find a subgraph with maximum minimum degree while satisfying a constraint on its diameter and being allowed to remove up to $k$ query vertices. We provide the following formal definition.

\smallskip

\begin{problem}[\mindeg] \label{prob:min_deg}
	Given a graph $G=(V_G,E_G)$, a set of query vertices $Q \subseteq V_G$, an integer $k \geq 0$ such that $k \leq |Q| -1$, an integer $diam_{\max} \geq 1$,   find an induced subgraph $H=(V_H,E_H)$ of $G$  such that:
	
	\begin{enumerate}
		\item $H$ is connected;
		\item $|V_H \cap Q| \geq |Q| - k$;
		\item $diam(H) \leq diam_{\max}$
                \item the minimum degree is maximized among all subgraphs satisfying 1-3;
	\end{enumerate}
\end{problem}

\smallskip

\begin{theorem}
	\mindeg\ (Problem~\ref{prob:min_deg}) is \NPhard\ even when $diam_{\max}=1$.
\end{theorem}

\smallskip

%
We omit the proof due to the space limit: we will present it in the extended version.
For Problem \ref {prob:min_deg} we provide a bicriteria algorithm (Algorithm \ref{alg:mindegree-diameter}, Theorem \ref{lemma:mainproc2}) in Section \ref{subsec:algo:mindegree-diameter}.
Finally, we also consider a variant where there is a constraint on the maximum distance between a query vertex and any other vertex in the solution subgraph $H$. We will show that a solution for such a problem can be found efficiently in polynomial time.

\begin{problem}[\mindegdist]\label{prob:min_deg2}
	Given a graph $G=(V_G,E_G)$, a set of query vertices $Q \subseteq V_G$, an integer $k \geq 0$ such that $k \leq |Q| -1$, an integer $ d_{\max} \geq1$,  find an induced subgraph $H=(V_H,E_H)$ of $G$  such that:
	
	\begin{enumerate}
		\item $H$ is connected;
		\item $|V_H \cap Q| \geq |Q| - k$;
		\item  for any $v \in V_H$, $d(V_H \cap Q,v) := \min_{q \in V_H \cap Q} d(q,v) \leq d_{\max} $
                \item the minimum degree is maximized among all subgraphs satisfying 1-3;
	\end{enumerate}
\end{problem}

For Problem \ref{prob:min_deg2} we provide a polynomial time exact algorithm (Algorithm \ref{alg:mindegree-distance}, Theorem \ref{theo:exact}) in Section \ref{subsec:algo:distance}.

\section{Algorithms and Analysis}
\label{sec:algorithms}
\vspace{-.5mm}

In this section we describe the algorithms for the three problems described in the previous section. \\
\spara{Minimizing the Diameter with Degree Constraint.}
We develop an algorithm for computing an approximate solution for \mindiam, where the objective is to minimize the diameter of the output graph while enforcing a lower bound on the minimum degree of the vertices. Our algorithm follows the framework developed in~\cite{SozioKDD10}, however, the variant with outliers has not been studied before, to the best of our knowledge.
The main procedure of our algorithm (Algorithm~\ref{alg:mindiammindeg}) receives a vertex $q \in Q$ in input and computes a feasible subgraph (if any) containing $q$ with ``small'' diameter. In particular, we shall show that if $q$ belongs to an optimum solution $O$ for \mindiam\ then our algorithm computes a $2$-approximation for such a problem. It proceeds as follows.  Let $G_0:=G$. At each step $t \geq 1$ the graph $G_t$ is obtained from $G_{t-1}$ as follows. If there is a vertex  violating the constraint on the degree, then such a vertex and all its edges are removed from $G_{t-1}$. Otherwise, a vertex at maximum distance from $q$ (and all its edges) is removed from $G_{t-1}$. Ties are broken arbitrarily.
The algorithm terminates as soon as the set of edges is empty or more than $k$ query vertices have been removed, at which step it computes the graph $H$ with minimum diameter among all $G_t$'s that are feasible. If none of the $G_t$'s is feasible, it returns ``unfeasible''. A pseudocode of our algorithm is shown in Algorithm~\ref{alg:mindiammindeg}.

\begin{algorithm}[ht]
\caption{\amindiam\ $(q)$} \label{alg:mindiammindeg}
\begin{algorithmic}[1]
\REQUIRE{$G=(V_G,E_G),Q \subseteq V$,$\delta_{\min}\geq 0$, $q \in Q$}
\ENSURE{ a subgraph $H\subseteq G$}
\STATE $G_0 \gets G$, $t \gets 0$
\WHILE{$E(V_t) \neq \emptyset$ and $|V_t \cap Q| \geq |Q| - k$ }
\STATE $t \gets t+1$
 \IF{there is $v$ in $G_{t-1}$ with $\delta(v) < \delta_{\min}$}
\STATE let $v$ be such a node
\STATE \textbf{If} $v=q$ \textbf{break}
\ELSE
\STATE let $v$ be a vertex with maximum $d_{G_{t-1}}(v,q)$
\ENDIF
\STATE $G_{t} \gets (V_{t-1} \setminus \{v\}, E(V_{t-1} \setminus \{v\})) $
\ENDWHILE

\STATE \textbf{If} there is no feasible graph among all $G_t$'s return \emph{unfeasible}
\STATE \textbf{else} return a graph with min max distance from $q$  among all feasible $G_t$'s.

\end{algorithmic}
\end{algorithm}
\setlength{\textfloatsep}{0pt}

The following lemma proves the approximation guarantees of our algorithm.

\smallskip

\begin{lemma}\label{lemma:mainproc}
If there is no feasible solution for \mindiam, Algorithm~\ref{alg:mindiammindeg} returns \emph{unfeasible}. Otherwise, let $O=(V_O,E_O)$ be an optimum solution for \mindiam. If $q \in V_O$,  then Algorithm~\ref{alg:mindiammindeg} computes a $2$-approximation solution for \mindiam.
\end{lemma}

\smallskip


We omit the proof due to the space limit.
The following theorem states our main result.

\smallskip

\begin{theorem}\label{thm:diam}
Running Algorithm~\ref{alg:mindiammindeg} for each of $k+1$ query vertices chosen arbitrarily, while outputting a feasible solution with minimum diameter computes a $2$-approximation algorithm for the \mindiam, while requiring $O(k \cdot  \max( |V_G|+|E_G|, |V_G| \log |V_G| ))$ operations when $\delta_{min} = 1$ and $O(|V_G| \cdot (|V_G|+|E_G|))$ operations in the general case.
\end{theorem}

\begin{IEEEproof}
The $2$-approximation guarantee follows from Lemma~\ref{lemma:mainproc} and from the fact that at least one of the $k+1$ query vertices considered by \amindiam$(q)$ must belong to an optimum solution (if any). To bound the number of operations required by the algorithm, observe that we need exactly one BFS for each of the $k+1$ query vertices in the case when there is no constraint on the minimum degree. This follows from the fact that vertices are deleted in non-increasing order of distance from $q$, which implies that the distance between any remaining vertex and $q$ is not affected. Therefore, the vertices can be sorted upfront according to their distance from $q$, requiring $O(k \cdot  \max( |V_G|+|E_G|, |V_G| \log |V_G| ))$ operations in this special case and $O(|V_G| \cdot (|V_G|+|E_G|))$ operations in the general case.
\end{IEEEproof}

Observe that the running time of our algorithm is almost linear (modulo a logarithmic factor) when there is no constraint on the minimum degree and $k$ is a constant. Otherwise, $O(|V_G| \cdot (|V_G|+|E_G|))$ operations are needed in the worst case, however, the actual number of operations required in real-world graphs are significantly smaller since we can remove in batch all the vertices $v$ with $\delta(v) < \delta_{\min}$.



\spara{Maximizing Minimum Degree  with Distance Constraint.}\label{subsec:algo:distance}
We develop an algorithm for computing an optimum solution to \mindegdist.  Let $G_0:=G$. At each step $t \geq 1$ the graph $G_t$ is obtained from $G_{t-1}$ as follows. For any vertex $v$ in $G_{t-1}:=(V_{t-1},E_{t-1})$, let $Q_v$ be the set of query vertices in the same connected component of $v$ in $G_{t-1}$. If there is a vertex $v$ such that  $|Q_v| < |Q| - k$ or $d_H(v,Q_v) := \min_{q \in Q_v} d_H(v,q) > d_{\max}$, then remove $v$ and all its edges. Otherwise, remove a vertex $v$ with minimum degree in $G_{t-1}$, breaking ties arbitrarily. The algorithm terminates when $G_{t}$ becomes empty, at which step it computes the graph $H$ with maximum minimum degree among all feasible connected components of all $G_t$'s. If there are several graphs $H$ with the same maximum minimum degree then it considers the graph $H$ with the largest number of query vertices. It then produces in output a connected component in $H$. If none of the $G_t$'s is feasible, it returns ``unfeasible''. A pseudocode of our algorithm is shown in Algorithm~\ref{alg:mindegree-distance}.
The following theorem proves the optimality of our algorithm.
\begin{algorithm}[ht]
\caption{\amindegdist} \label{alg:mindegree-distance}
\begin{algorithmic}[1]
\REQUIRE{$G=(V_G,E_G),Q \subseteq V$, $d_{\max}>0$}
\ENSURE{ a subgraph $H$ of $G$}
\STATE $G_0 \gets G$, $t \gets 0$
\WHILE{$E(V_t) \neq \emptyset$}
\STATE $t \gets t+1$
\STATE Let $Q_v$ be the set of query vertices in the same connected component of $v$ in $G_{t-1}$.
 \IF{there is a vertex $v$ such that $|Q_v| < |Q| - k$ or $d_{G_{t-1}}(v,Q_v) > d_{\max}$}
\STATE let $v$ be such a node
\ELSE
\STATE let $v$ be a vertex with minimum degree in $G_t$
\ENDIF
\STATE $G_{t} \gets (V_{t-1} \setminus \{v\}, E(V_{t-1} \setminus \{v\})) $
\ENDWHILE

\STATE \textbf{If} there is no feasible graph among all $G_t$'s return \emph{unfeasible}
\STATE \textbf{else} return a subgraph with maximum minimum degree among all feasible connected components of all $G_t$'s
\end{algorithmic}
\end{algorithm}
\setlength{\textfloatsep}{0pt}

\smallskip

\begin{theorem}\label{theo:exact}
If there is no feasible solution for \mindegdist, Algorithm~\ref{alg:mindegree-distance}  returns unfeasible. Otherwise it computes an optimum solution for \mindegdist.
\end{theorem}

\begin{IEEEproof}
If there is no feasible solution then the algorithms outputs unfeasible, in that, the algorithm makes sure that only feasible solutions are produced in output. If there is a feasible solution, then there must be a smallest step $t$ such that there is $v \in O$ with $v \in V_t \setminus V_{t+1}$, as the algorithm eventually removes all vertices in the graph. Let $Q_v$ be the set of query vertices in the same connected component of $v$ in $G_t$. It holds that a) $|Q_v| \geq |Q| - k$ and b) $d_{G_t}(v,Q_v) \leq d_{\max}$. This follows from the facts that $O$ is an induced subgraph of $G_t$, $d_O(u,v) \geq d_{G_t}(u,v)$ for all $u,v$ in $O$, and any connected component in $O$ is also connected in $G_t$. As a result, every vertex in $G_t$ satisfies  a) and b), for otherwise the algorithm would have removed a vertex violating at least one such a constraint.
Moreover, $v$ has minimum degree in $G_t$. Let $C$ be the connected component in $G_t$ containing $O$. The following chain of inequalities proves the optimality of $C$:
\[
\delta^{\min}(O) \leq \delta_O(v) \leq  \delta_{C}(v) = \delta^{\min}(C),
\]
where the first inequality follows from $v \in O$, the second inequality follows from the fact that $O$ is contained in $C$, while the equality follows from the way $v$ is chosen.
\end{IEEEproof}

\spara{Implementation Details.} Our algorithm requires to compute a BFS at each step, which might be cumbersome in the case when the input graph is large. Such a problem can be alleviated by a preprocessing step which proceeds as follows. First, we run a BFS starting from each of the query vertices and compute the minimum distance between every vertex in the input graph and the query vertices. Then, all vertices violating the distance constraints are removed all together. It is easy to see that none of those vertices can be part of any optimum solution. We shall refer to such a preprocessing step as \emph{pruning}. After that, Algorithm~\ref{alg:mindegree-distance} is executed in the resulting graph. In most cases of interest, the pruning phase allows to filter out a large fraction of vertices in the input graph, allowing Algorithm~\ref{alg:mindegree-distance} to be executed efficiently. Let $H=(V_H,E_H)$ be the graph obtained after the pruning phase. The running time of the pruning phase is $O( |Q| \cdot (|V_G| + |E_G|)  )$, while the running time of our algorithm after pruning is $O(|Q| \cdot (|V_H+E_H|) \cdot |V_H| )$. Our experiments on real-world data show that our algorithm is much more efficient in practice than what the worst-case analysis suggests.

\spara{Maximizing Minimum Degree  with Diameter Constraint.}
\label{subsec:algo:mindegree-diameter}

We develop an algorithm for \mindeg\ where the objective is to maximize the minimum degree subject to a constraint on the diameter of the graph. The pseudocode of the main procedure is shown in Algorithm~\ref{alg:mindegree-diameter}.
\begin{algorithm}[ht]
\caption{\textsc{\amindeg}$(q)$} \label{alg:mindegree-diameter}
\begin{algorithmic}[1]
\REQUIRE{$G=(V_G,E_G),Q \subseteq V$,$diam_{\max}\geq 1$, $q \in Q$}
\ENSURE{ a subgraph $H$ of $G$}
\STATE $G_0 \gets G$, $t \gets 0$
\WHILE{$E(V_t) \neq \emptyset$ and $|V_t \cap Q| \geq |Q| - k$ }
\STATE $t \gets t+1$
 \IF{there is $v$ in $G_{t-1}$ with $d(q,v) >  diam_{max}$}
\STATE let $v$ be such a node
\ELSE
\STATE let $v$ be a vertex with minimum degree in $G_{t-1}$
\ENDIF
\STATE \textbf{if} $v=q$ \textbf{break}
\STATE $G_{t} \gets (V_{t-1} \setminus \{v\}, E(V_{t-1} \setminus \{v\})) $
\ENDWHILE

\STATE \textbf{If} there is no feasible graph among all $G_t$'s return \emph{unfeasible}
\STATE \textbf{else} return a graph with max min degree among all feasible $G_t$'s.

\end{algorithmic}
\end{algorithm}
\setlength{\textfloatsep}{0pt}

Similarly to the algorithm for \mindeg, we run Algorithm~\ref{alg:mindegree-diameter} for each of $k+1$ query vertices (chosen arbitrarily) and output the solution with max min degree. The following theorem states the approximation guarantees of our algorithm.

\smallskip

\begin{theorem}\label{lemma:mainproc2}
If there is no feasible solution for \mindeg, Algorithm~\ref{alg:mindegree-diameter} returns \emph{unfeasible}. Otherwise, let $O=(V_O,E_O)$ be an optimum solution for \mindeg\ and let $H$ be the solution computed by our algorithm for \mindeg. It holds that 1) $\delta_{\min}(H) \geq \delta_{min}(O)$, and 2) $diam(H) \leq 2 \cdot diam (O)$.
\end{theorem}

\smallskip 

The proof is similar to the proof of Theorem~\ref{thm:diam} and it is omitted for space constraints. Our algorithm can be seen as a bicriteria algorithm (see for example~\cite{DBLP:journals/siamcomp/GrandoniKPS08}). That is, an algorithm for a problem with multiple objective functions, which in our case are the minimum degree and the diameter of the output graph.
The worst case complexity  is $O(k\cdot |V_G| \cdot (|V_G|+|E_G|))$ since, for each deleted vertex we should recompute all the distances from $q$, for each query vertex $q$. However, if $max\{d(q,v)\} >  diam_{max}$, we can speed up the computation removing in batch all the vertices $v$ such that  $d(q,v) >  diam_{max}$.

%
%
%
\section{Experiments}
\label{sec:experiments}
\vspace{-1mm}
In this section we assess the efficiency and the effectiveness of the three community search with outliers methods. In particular we focus our analysis on the following main aspects:
\begin{enumerate}
  \item the capability of \mindeg, \mindegdist~and  \mindiam~to detect query vertices that are outliers;
  \item comparison with the Minimum Inefficiency Subgraph proposed in~\cite{RuchanskyBGGK17} in the outlier classification task;
  \item characteristics of the solution subgraphs produced, varying the parameter $k$, i.e., the number of query vertices that we are allowed to drop, and varying the constraints;
  \item characteristics of the solution subgraphs produced varying the query set $Q$ w.r.t. its size, and the distance among its vertices;
  \item runtime.
\end{enumerate}

For our purposes we use four real-world networks described in Table~\ref{tab:graphs}.
\setlength{\textfloatsep}{0pt}
\begin{table}[t!]
	\scriptsize
	
	\centering
	\caption{Characteristics of datasets used.}
	
	\label{tab:graphs}
	
		\vspace{-1.5mm}
	\begin{tabular}{r|c|c|c|c|c|}
		
		\multicolumn{1}{c}{}	 & \multicolumn{1}{c}{$|V|$} & \multicolumn{1}{c}{$|E|$} &  \multicolumn{1}{c}{density}&  \multicolumn{1}{c}{avg deg}   & \multicolumn{1}{c}{diam.} \\
		\cline{2-6}
		\texttt{amazon} & 334,863 &925,872  &1.6e-5&5.5&44 \\
		\texttt{dblp}&317,080&1,049,866&2.1e-5&6.62&23\\
		\texttt{youtube}&1,138,499&2,990,443&4.6e-6&5.27&21\\
		\texttt{ljournal}&3,997,962&34,681,189&4.3e-6&17.3&16\\
		
		\cline{2-6}
	\end{tabular}
	
\end{table}

All datasets are publicly-available, undirected and unweighted:
\texttt{youtube}\footnote{\label{snap}\url{http://snap.stanford.edu}} and \texttt{ljournal}~\textsuperscript{\ref{snap}} are social networks, \texttt{dblp}~\textsuperscript{\ref{snap}}  is a collaboration network and \texttt{amazon}~\textsuperscript{\ref{snap}} is a co-purchasing network.
All our datasets, come with auxiliary \emph{ground-truth communities} information.
This information is needed to create query workloads with known number of outliers (i.e., query vertices coming from a different community w.r.t. the majority of  the other query vertices), as it will be described in details later. The largest dataset, \texttt{ljournal}, is used only to report runtime.
All algorithms are implemented in \texttt{C++}, building on the open-source NetworKit framework\footnote{\url{http://networkit.iti.kit.edu/}}.
The experiments are conducted on a server equipped with 32 GB RAM and with a processor Intel 1.2 GHz CPU.

\begin{table}[t!]
		\scriptsize
	\vspace{2mm}
	\caption{ Outliers identification capabilities of the three algorithms on three datasets, with $n=10$ and $m = k \in [1,4]$. The table reports the average number of outliers detected (and the percentage w.r.t. the total number of outliers present). A dash reports an experiment which does not have a feasible solution in at least 50\% of the runs.  \label{tab:outlier_identification}}
	\setlength\tabcolsep{2pt}
	\vspace{-1.5mm}
	\begin{tabular}{|c|c|c|c|c|c|}
			\multicolumn{1}{c}{}  & \multicolumn{1}{c}{Problem}    &  \multicolumn{1}{c}{} & \multicolumn{1}{c}{} &\multicolumn{1}{c}{} &\multicolumn{1}{c}{}  \\
				\multicolumn{1}{c}{Dataset}  & \multicolumn{1}{c}{[Constraint]}      &  \multicolumn{1}{c}{$k=1$} & \multicolumn{1}{c}{$k=2$} &\multicolumn{1}{c}{$k=3$} &\multicolumn{1}{c}{$k=4$}  \\
			
		\hline
		\texttt{dblp}    &       & 1  (100\%)& 1.9 (95\%)  &2.8  (93.3\%)  &3.8  (99\%) \\
		\texttt{amazon}  &    \texttt{\mindeg}     & 0.9 (90\%)   & 1.9   (95\%)  &2.9  (96.7\%)    &3.7  (97.9\%)   \\
		\texttt{youtube} &       [$diam_{max}=4$]   & 0.6 (69\%) & 1.3 (65\%) & 1.6 (53.3\%)  & 3.73 (93\%) \\
		\cline{1-6}
		
		\texttt{dblp}    &      & 0.5 (50\%) &1.3  (65\%)  &1.8  (60\%)  & 2.7 (67.5\%)  \\
		\texttt{amazon}  &  \texttt{{\mindegdist}} &  0.9(90\%)    &1.9  (95\%)     &2.9 (96.7\%)    &3.9 (97.5\%)  \\
		\texttt{youtube} &  [$d_{max}=2$]  & 0.4 (40\%) & 0.6 (30\%)  & 0.9 (30\%) & 1.4 (35\%)  \\
		
		\cline{1-6}
		\texttt{dblp}    &        &0.3 (30\%)     &0.8 (40\%)    &1.2  (40\%)  &  1.8(45\%)  \\
		\texttt{amazon}  &  \texttt{\mindiam}      & -& 0.7(35\%)  & 0.9 (30\%)  & 1.4 (35\%)  \\
		\texttt{youtube} &    [$\delta_{min}=3$]     &0.5(90\%)    & 1.0 (50\%)  & 1.5 (50\%) & 2.0 (50\%) \\
		\hline
	\end{tabular}
	
\vspace{2mm}
\end{table}

\begin{figure}[t!]
	\begin{tabular}{cc}	\hspace{-6mm}\includegraphics[width=.28\textwidth]{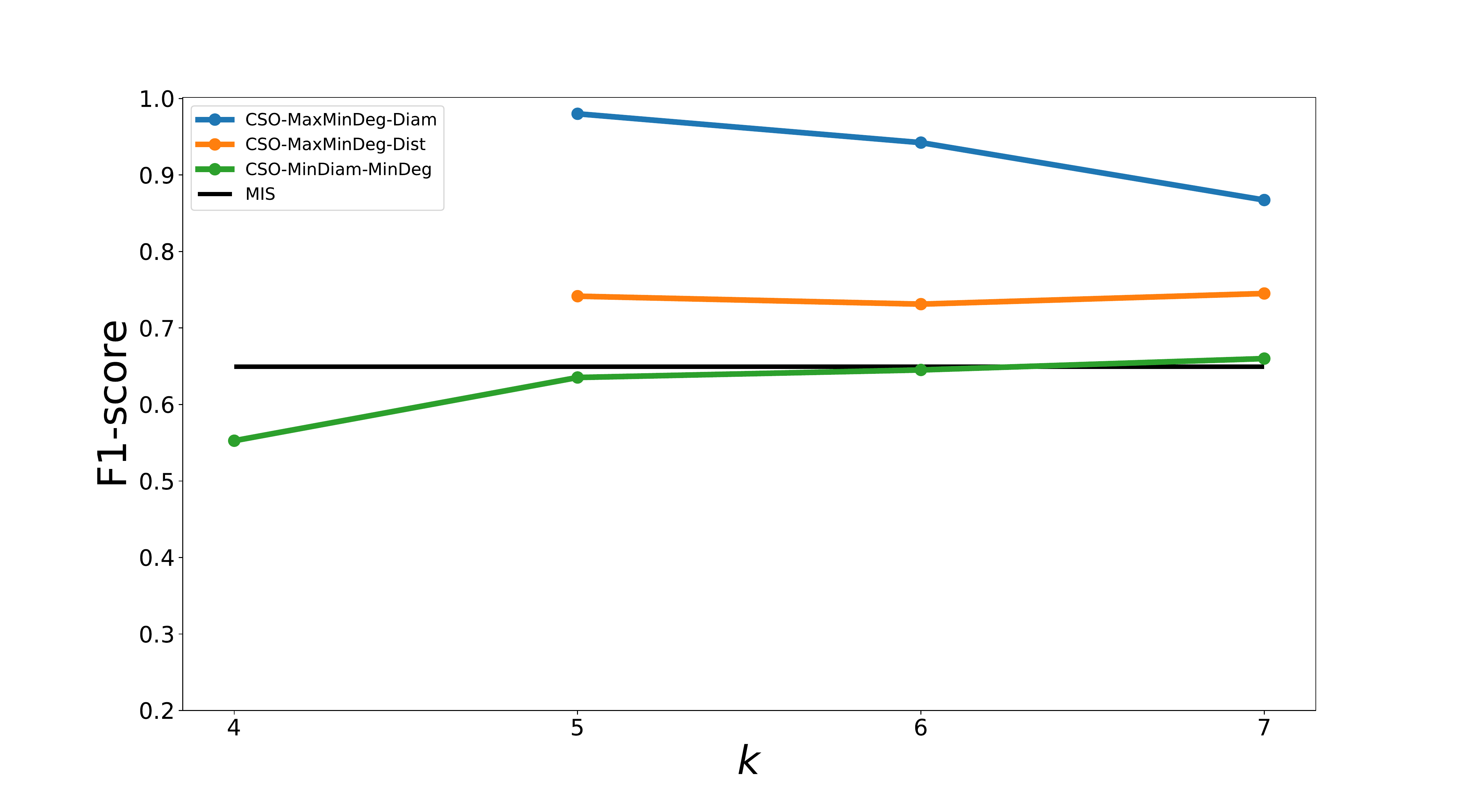}& \hspace{-8mm}\includegraphics[width=.28\textwidth]{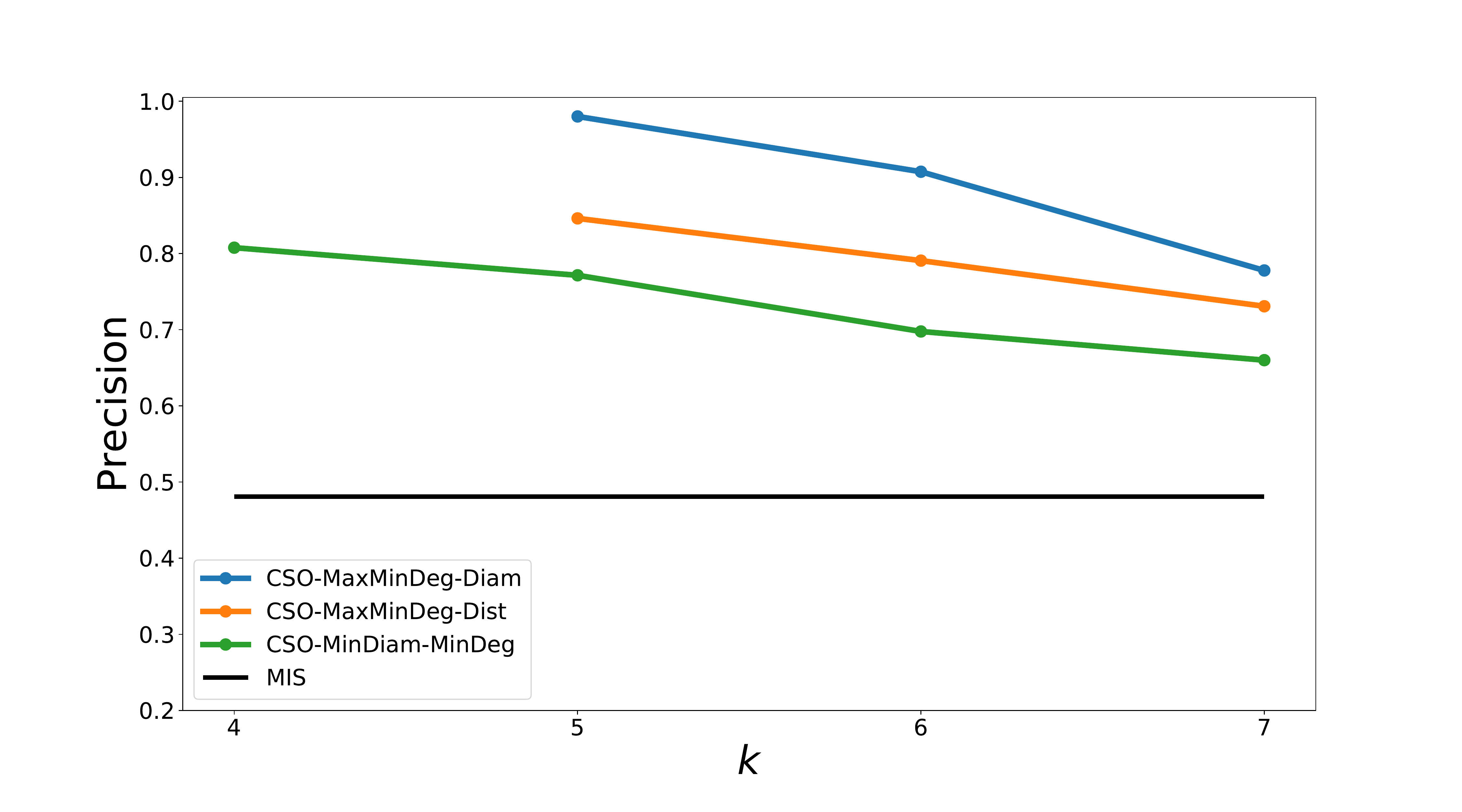}\\
		\hspace{-6mm}\includegraphics[width=.28\textwidth]{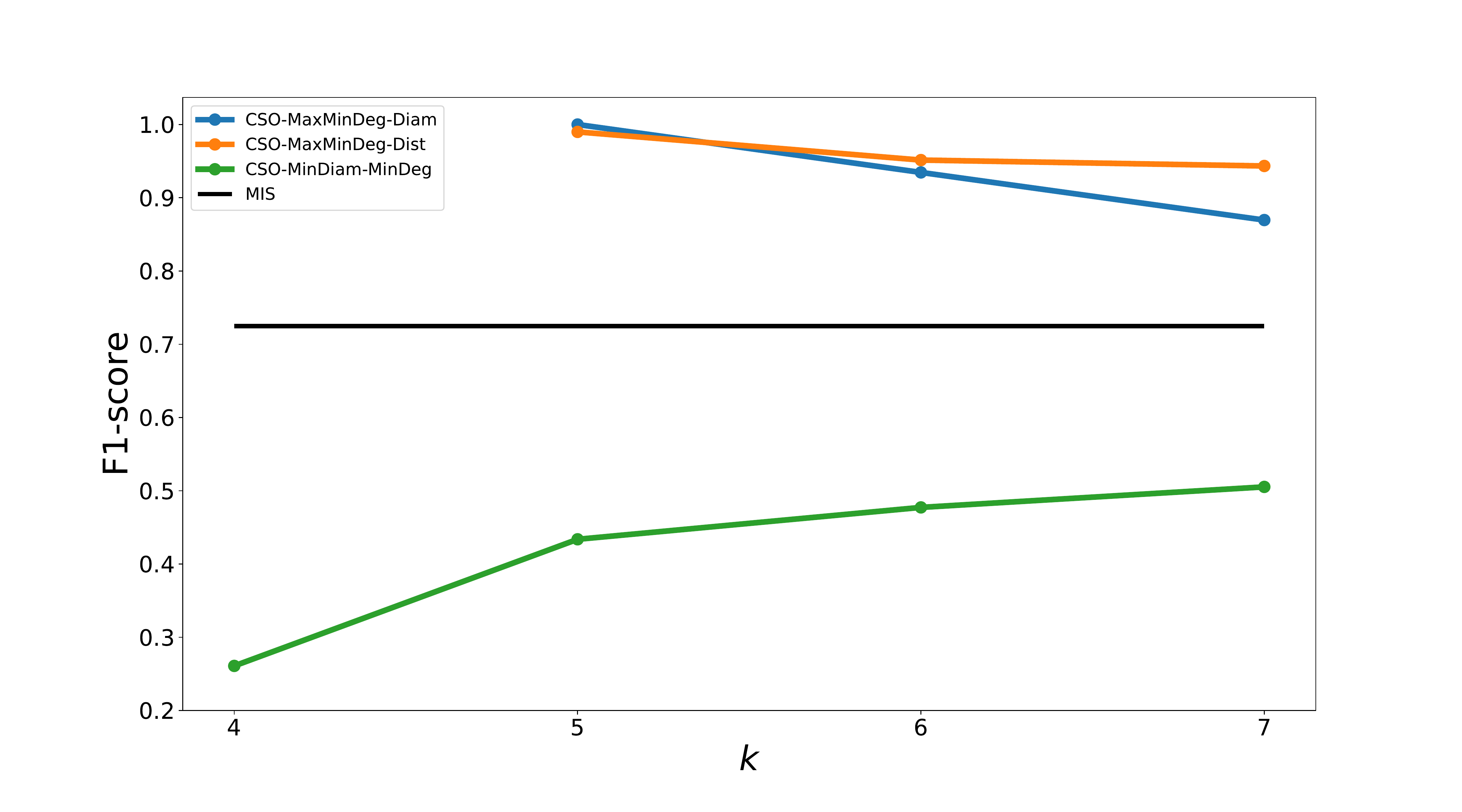}& \hspace{-8mm}\includegraphics[width=.28\textwidth]{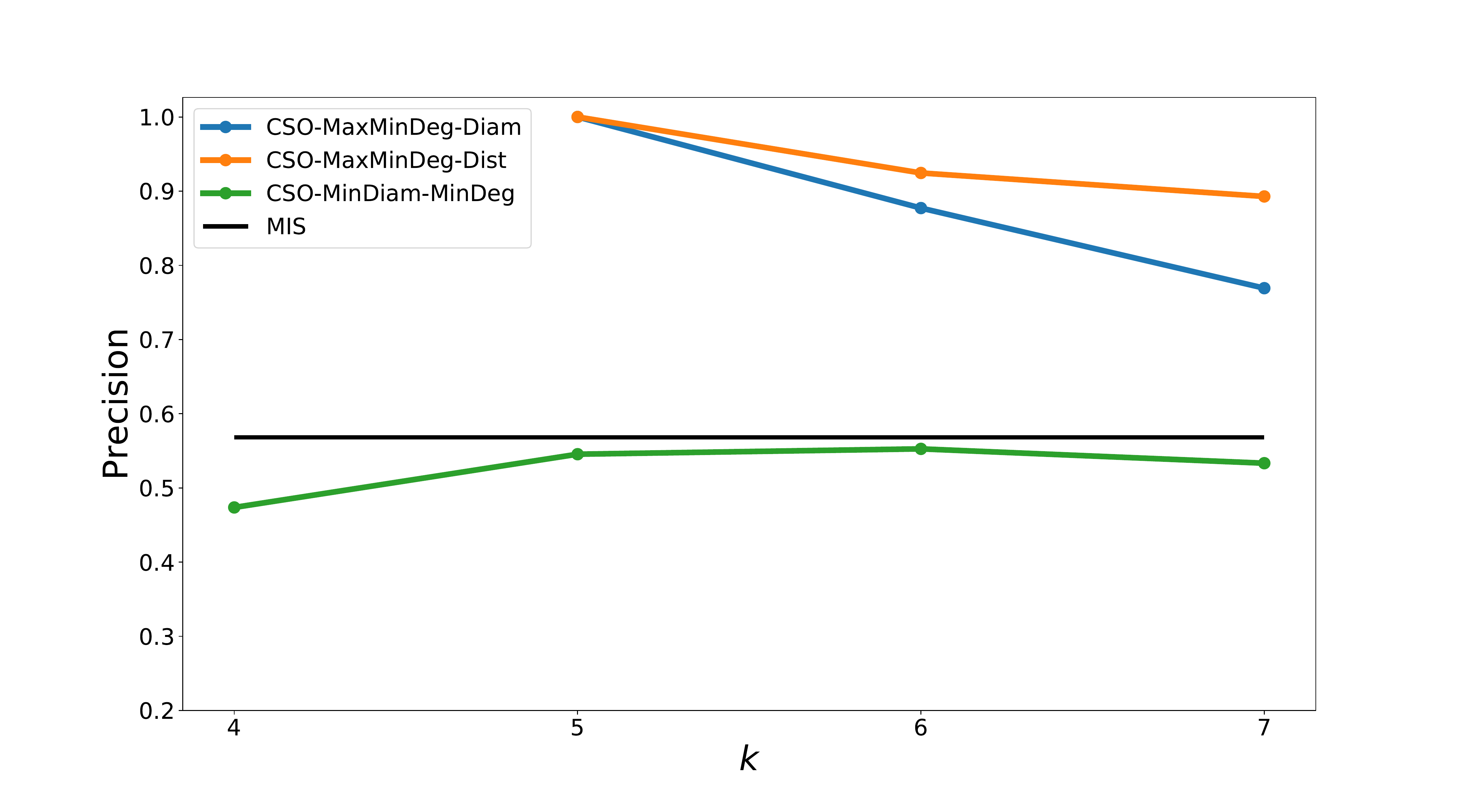}\\
	\end{tabular}
\vspace{-2mm}
	\caption{Average F1-Score and average Precision of the solutions on \texttt{dblp} (top) and \texttt{amazon} (bottom) for \mindiam~($diam_{max}=3$), \mindegdist~($d_{max}=2$), \mindeg~($\delta_{min}=3$) and MIS.}
	\label{fig:mis}
\vspace{2mm}
\end{figure}



\smallskip

\spara{Capability to detect outliers.}
We first study the capability of \mindeg, \mindegdist~and  \mindiam~to detect outliers. We randomly select ten vertices (i.e $n=10$) inside the same community (randomly chosen). Then, select $m$ vertices at random from different communities w.r.t. the one from which the first $n$ query vertices have been selected.
As these communities are ground-truth communities, we can consider the vertices coming from different communities as unrelated, and if there is a dominating group of vertices (the $n$ vertices chosen from a unique community) then the others can be considered ground-truth outliers that we can control. In this experiment vary $m$ in the range $[1,4]$ and we set $m=k$: in this way, the number of effective outliers (i.e. $m$) is equal to the number of the query vertices that we are allowed to drop (i.e. $k$).
We generate 20 query sets following the process just described and we report the average value.
Since some query sets may not have any solution for some problem and for some constraint, we report the value only the first 10 sets return a solution.
Table \ref{tab:outlier_identification} reports the average number of outliers detected (and the percentage w.r.t. the total number of outliers present) for the three algorithms under three datasets. We can observe how in general, all the algorithms perform quite well in detecting the outliers injected in the query sets, with values ranging  in between 30\% and 100\%.

\spara{Outliers classification task evaluation.}
We next compare against the Minimum Inefficiency Subgraph (MIS) studied in~\cite{RuchanskyBGGK17}.
It is important to note that the problem in~\cite{RuchanskyBGGK17} is totally parameterless, so that it decides automatically the number of outliers. In particular, the formulation of ~\cite{RuchanskyBGGK17} always requires all the query vertices to be part of the solution, but on the other hand, it allows the solution subgraph to be \emph{disconnected}. This way, query vertices which are not in the largest connected component (LCC) are considered outliers.
We  select $n=10$ query vertices inside the same community (with maximum distance among them equal to 4) and $m=5$ outliers. We run the Minimum Inefficiency Subgraph and \mindiam~($diam_{max}=3$), \mindegdist~($d_{max}=2$), \mindeg~($\delta_{min}=3$) on \texttt{dblp}  and 	\texttt{amazon} varying the parameter $k\in [4,7]$ . We repeat the experiment 10 times.
From every MIS solution,  we extract the largest connected component (LCC) to identify the query vertices which \emph{are not} in the LCC and mark them as outliers.
We compute the average F1-score and the average precision varying the parameter $k$ and we report the results in Figure~\ref{fig:mis}. Since MIS does not require any parameter in input, its score is constant for every $k$.
We notice that, for $k\geq 5$, \mindiam~and \mindegdist~achieve a better F1-score and a better precision on both	\texttt{dblp}  and 	\texttt{amazon}. On the other hand, \mindeg ~ has worst performance since it returns larger solutions than the other methods due to the fact that \texttt{amazon} and \texttt{dblp} networks have a low average degree.
For $k=4$  \mindiam~and  \mindegdist~does not return any solution for at least one query set of vertices: however in this case it is easy to see that the average F1-score is greater or equal to 0.5 since every vertex is classified as outlier.


\spara{Characterization varying constraints and $k$.}
We next study how the solutions change varying the number of allowed outlier vertices $k$.
In this set of experiments we no longer set $k$ to the known number of outliers in the query set.
The query workload generation is as follows.
We randomly pick a community and we select randomly $n = 10$ vertices from the community,
 with distance at most 4 from each other. Then we select $m= 4$ outliers randomly in the whole network. As before, we generate 20 query sets and we report the average value of the first 10 query sets.
 In particular we report three measures:
  \begin{enumerate}
    \item the number $Q_H$ of query vertices contained in the solution subgraph $H$;
    \item the relative size of $H$ (i.e the ratio between the number of vertices in $H$ ($|H|=|V_H|$) divided the number of vertices in $G$ ($|G|=|V_G|$);
    \item the minimum degree within $H$.
  \end{enumerate}

\begin{figure}
	\begin{tabular}{cc}		
				\hspace{-4mm}\includegraphics[width=.25\textwidth]{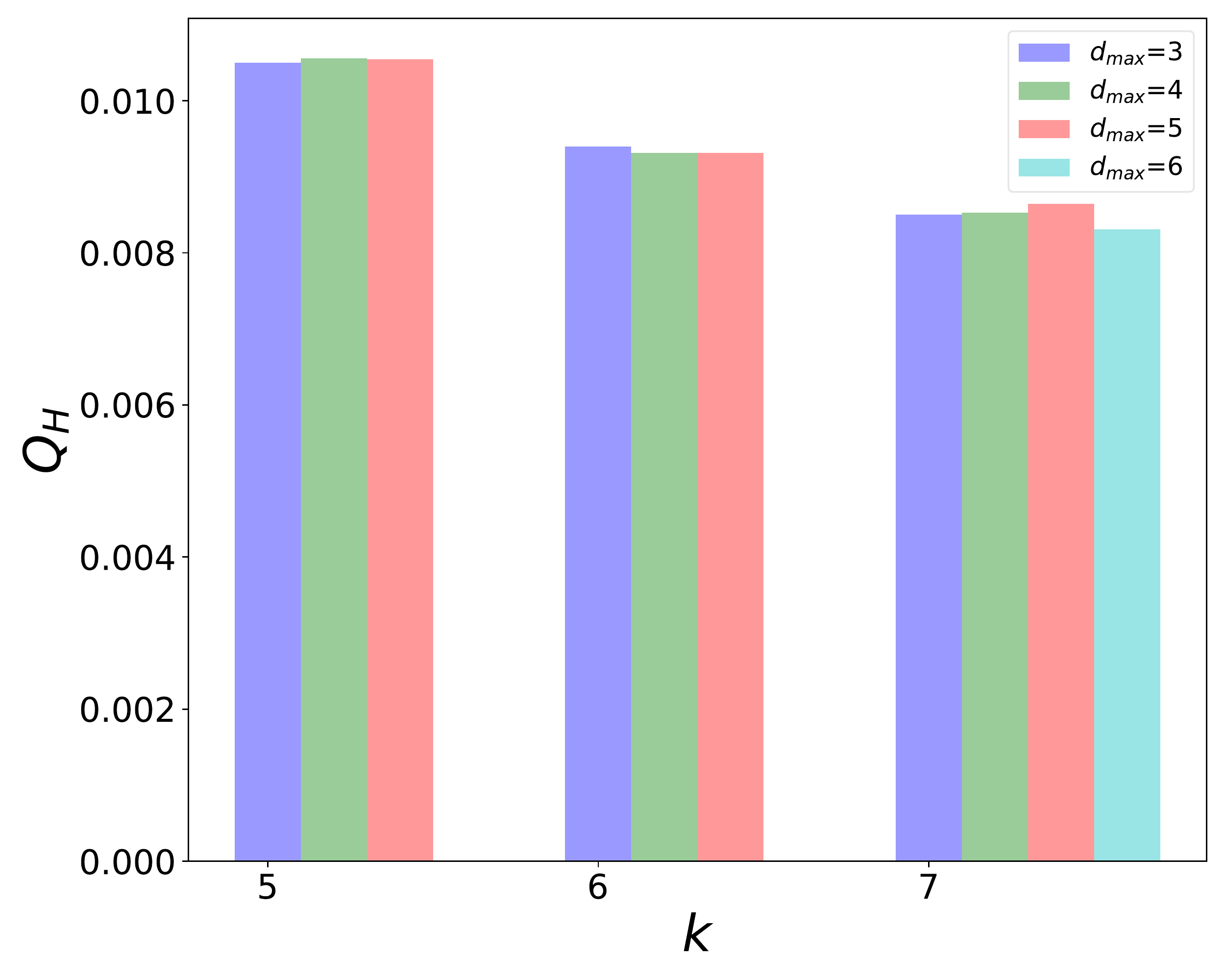}&	\hspace{-4mm}\includegraphics[width=.25\textwidth]{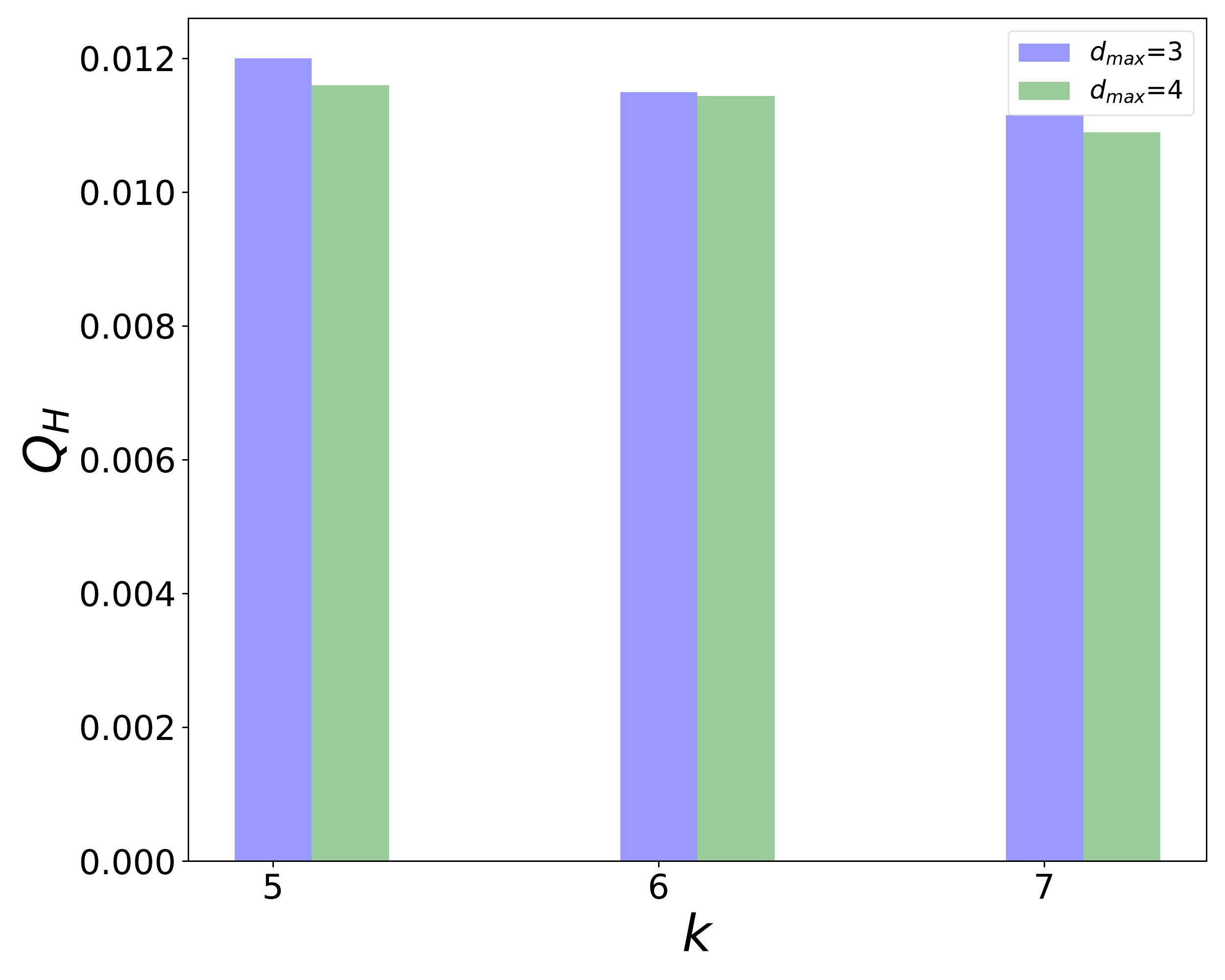}\\
		\hspace{-4mm}\includegraphics[width=.25\textwidth]{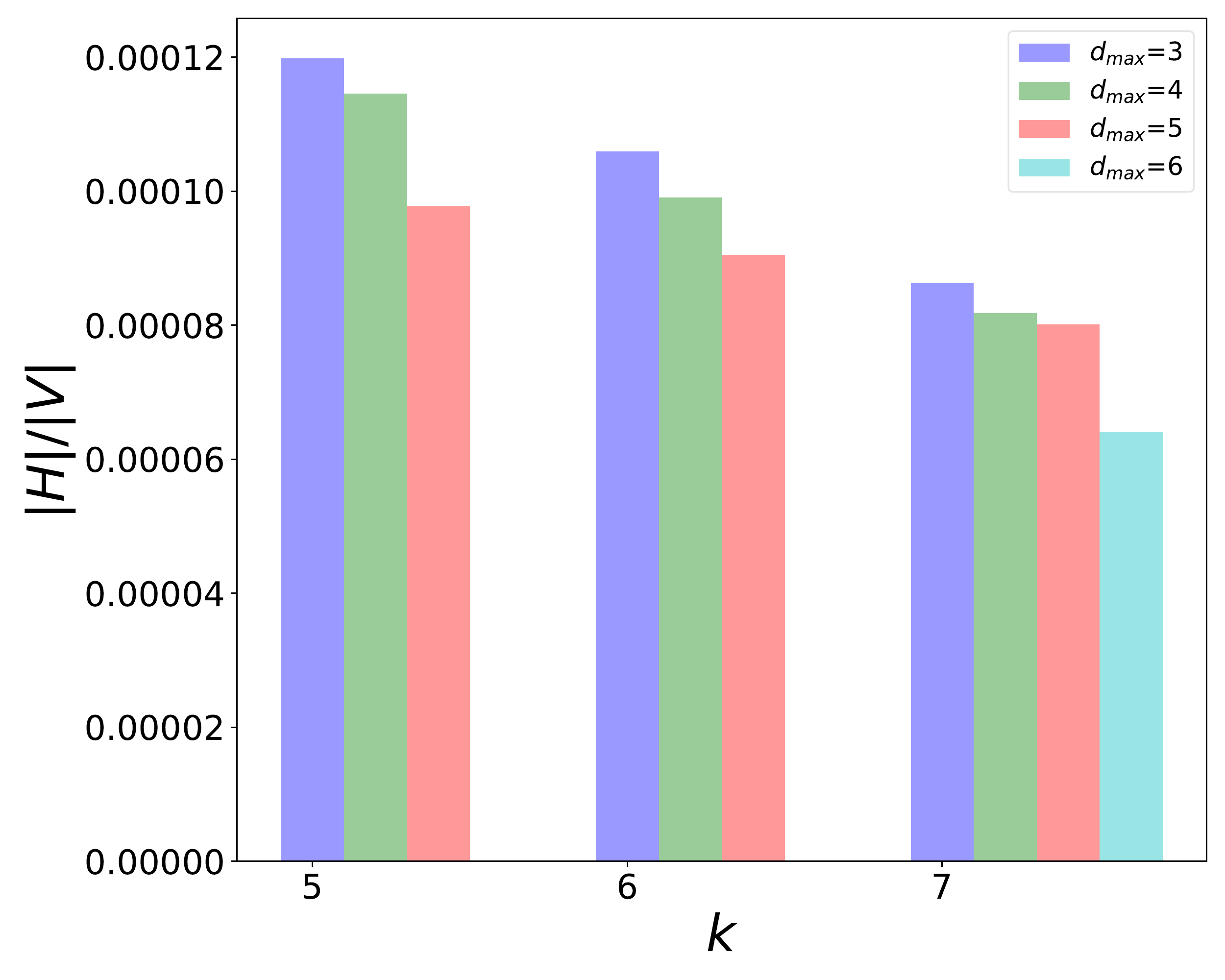}&	\hspace{-4mm}\includegraphics[width=.25\textwidth]{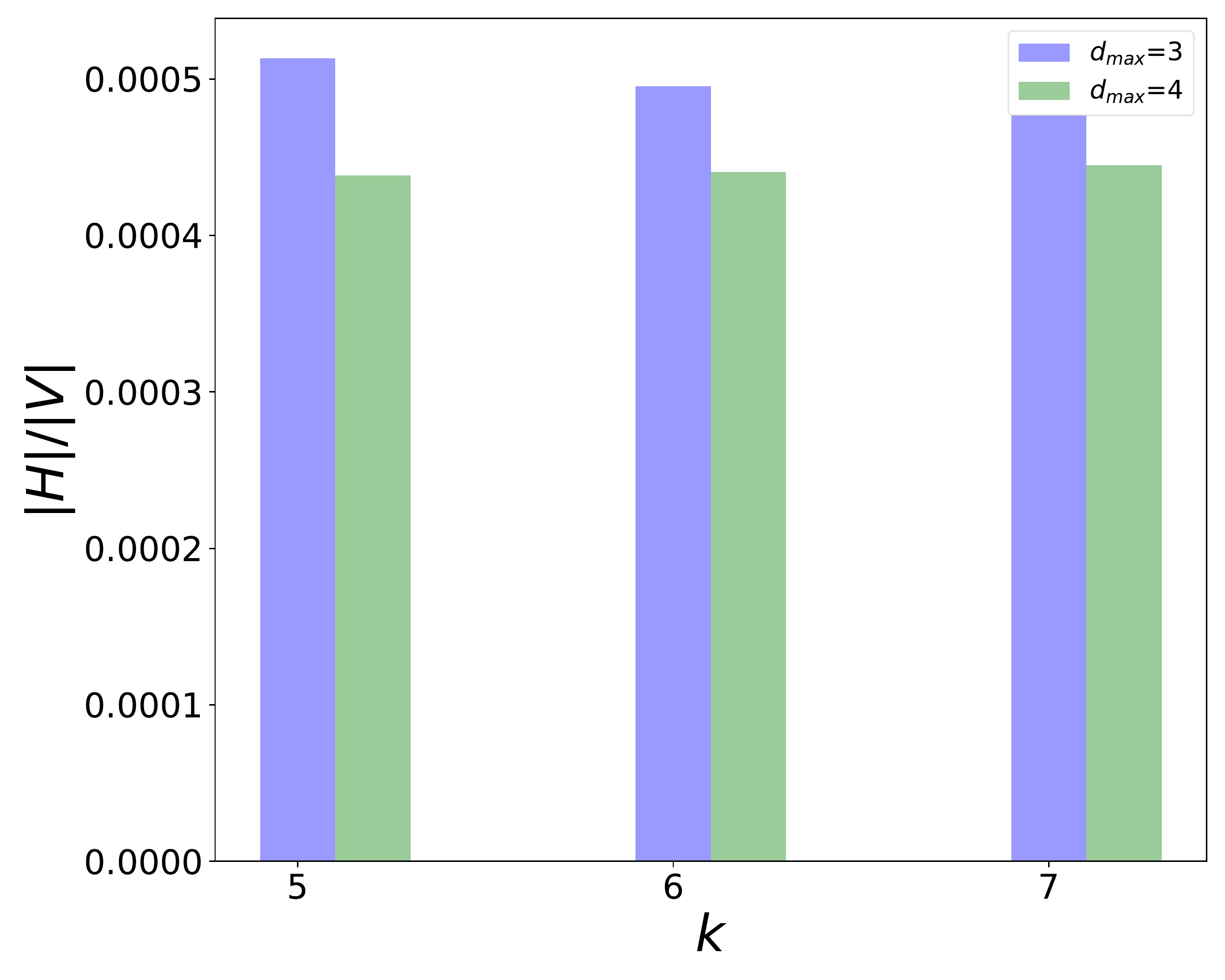}\\
						\hspace{-4mm}\includegraphics[width=.25\textwidth]{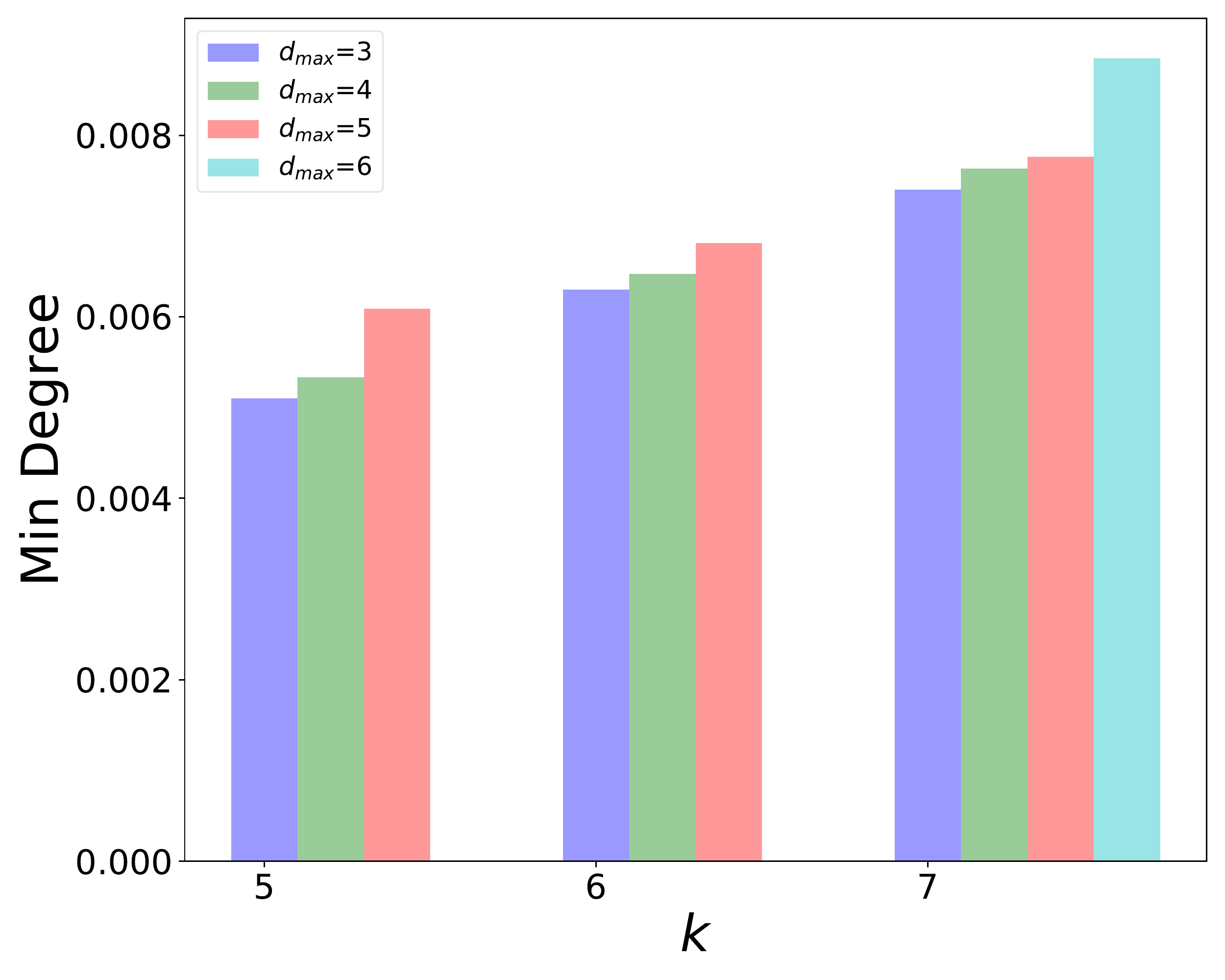}&	\hspace{-4mm}\includegraphics[width=.25\textwidth]{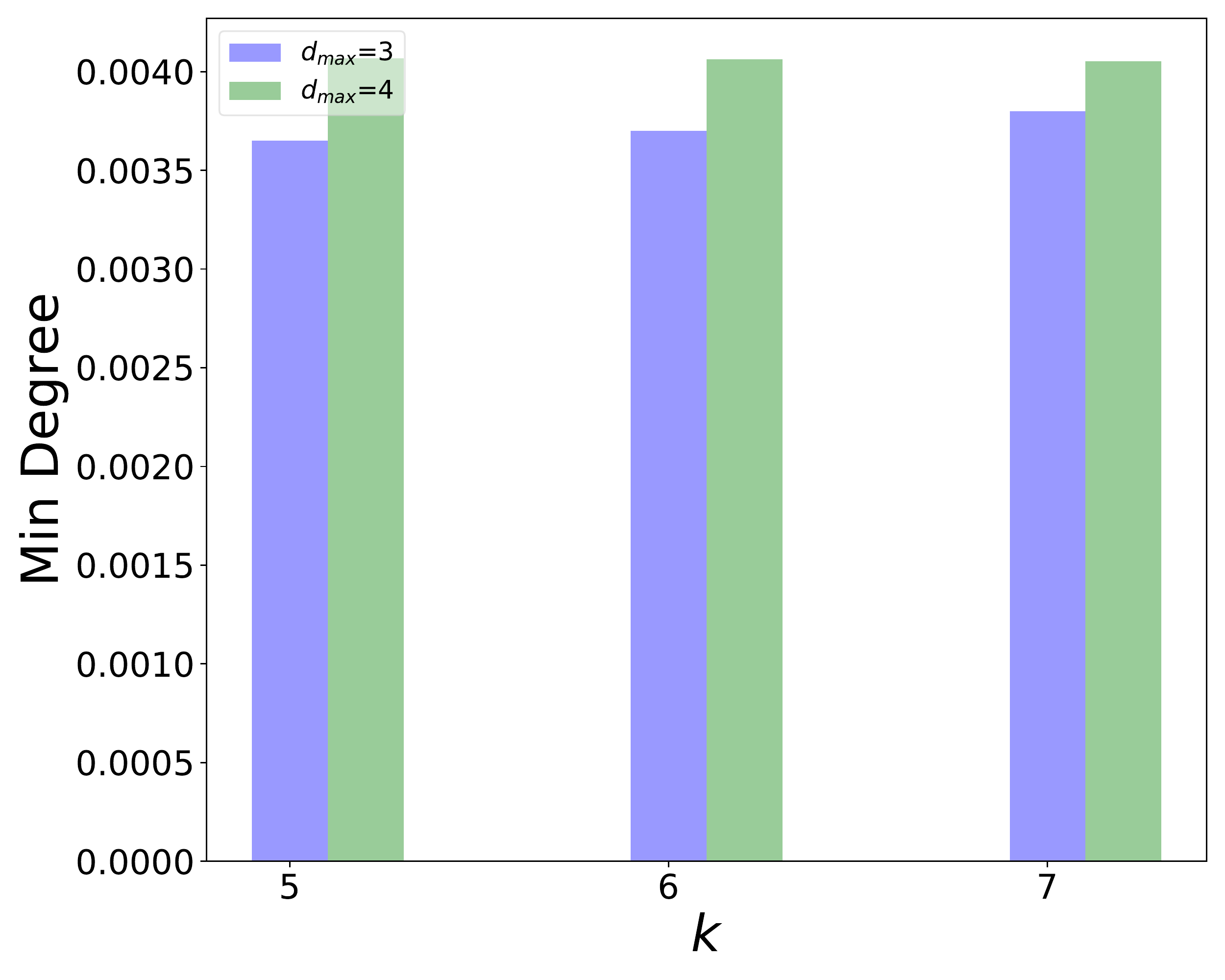}\\
	\end{tabular}
\vspace{-2mm}
		\caption{Characteristics of the solution on \texttt{youtube} (left) and \texttt{amazon} (right)  for \mindiam~with $k \in \{4,6,8\}$ and the minimum degree threshold $\delta_{\min}$ in the range $[3,6]$.}
	\label{fig:dblp-youtube-mindiam}
\vspace{2mm}
\end{figure}

\begin{figure}[t!]
	\begin{tabular}{cc}	
		\hspace{-4mm}\includegraphics[width=.25\textwidth]{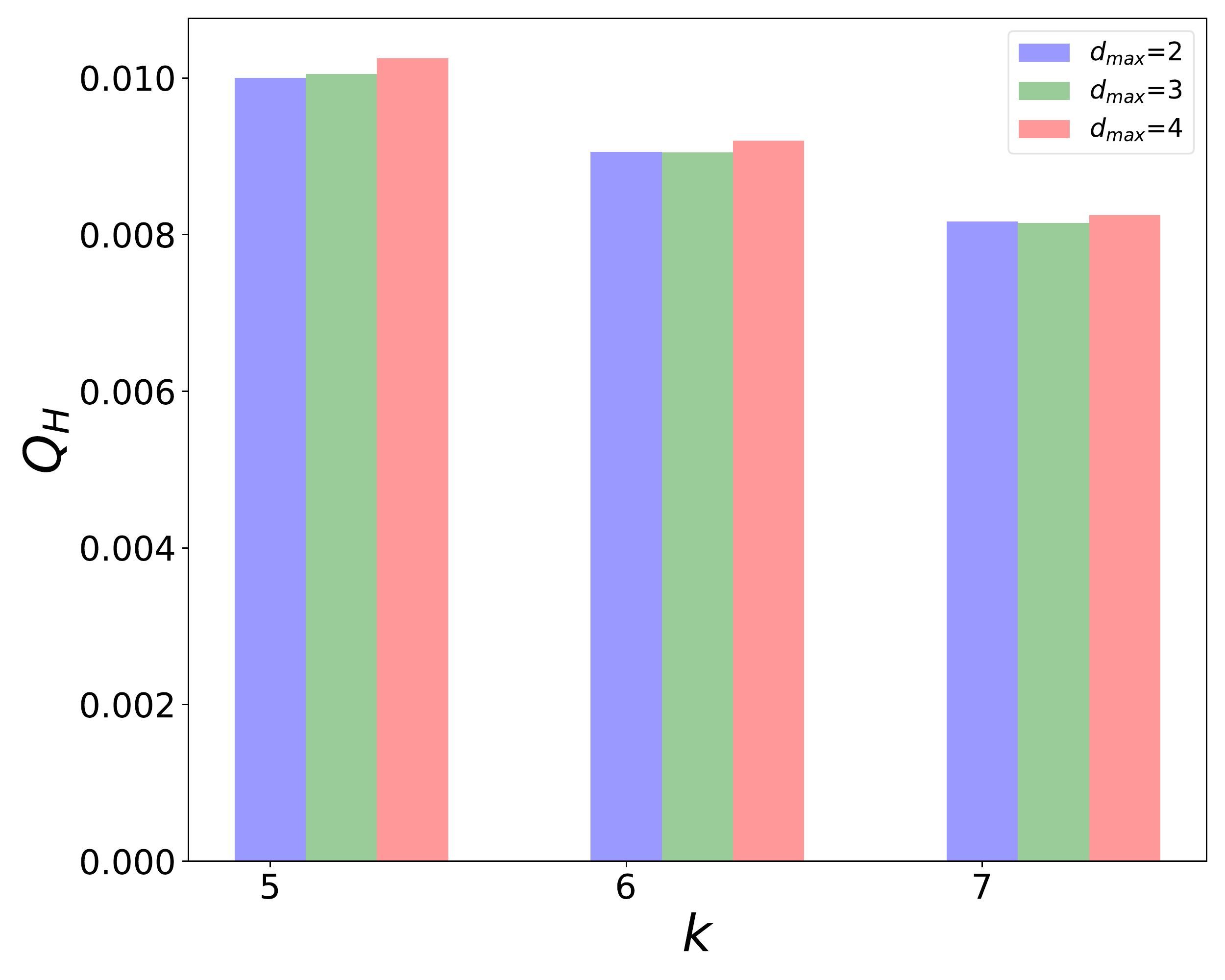}&	\hspace{-4mm}\includegraphics[width=.25\textwidth]{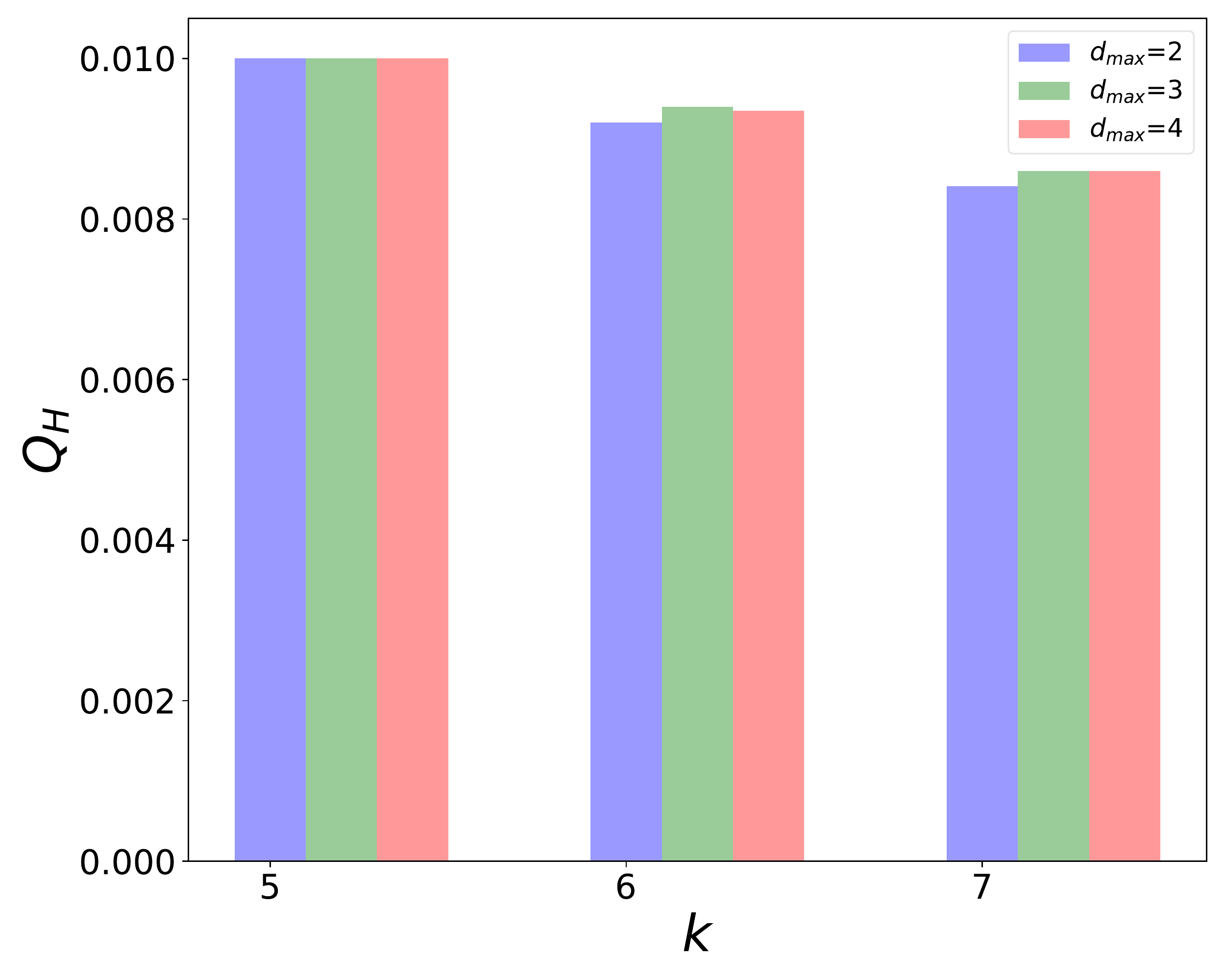}\\
		\hspace{-4mm}\includegraphics[width=.25\textwidth]{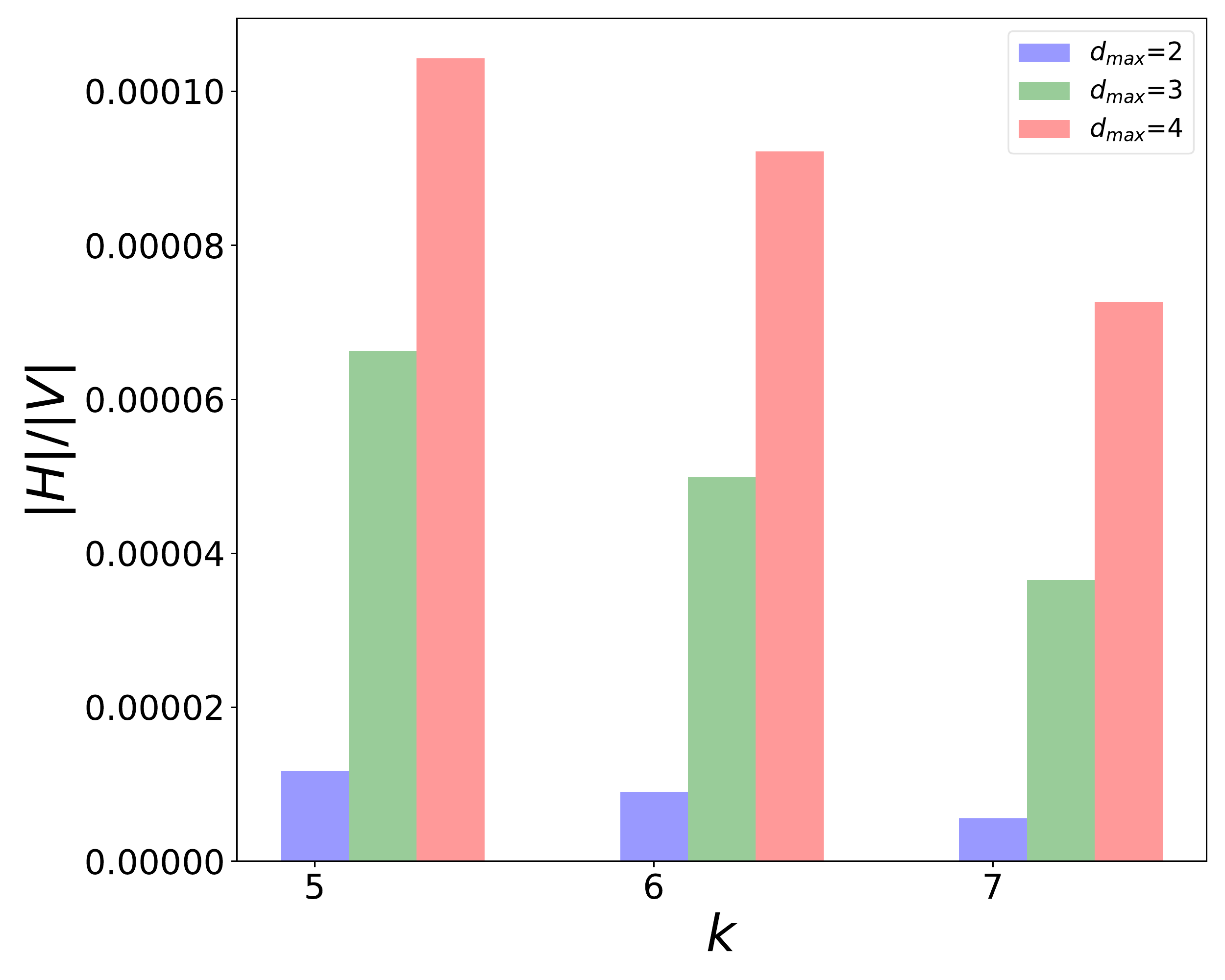}&	\hspace{-4mm}\includegraphics[width=.25\textwidth]{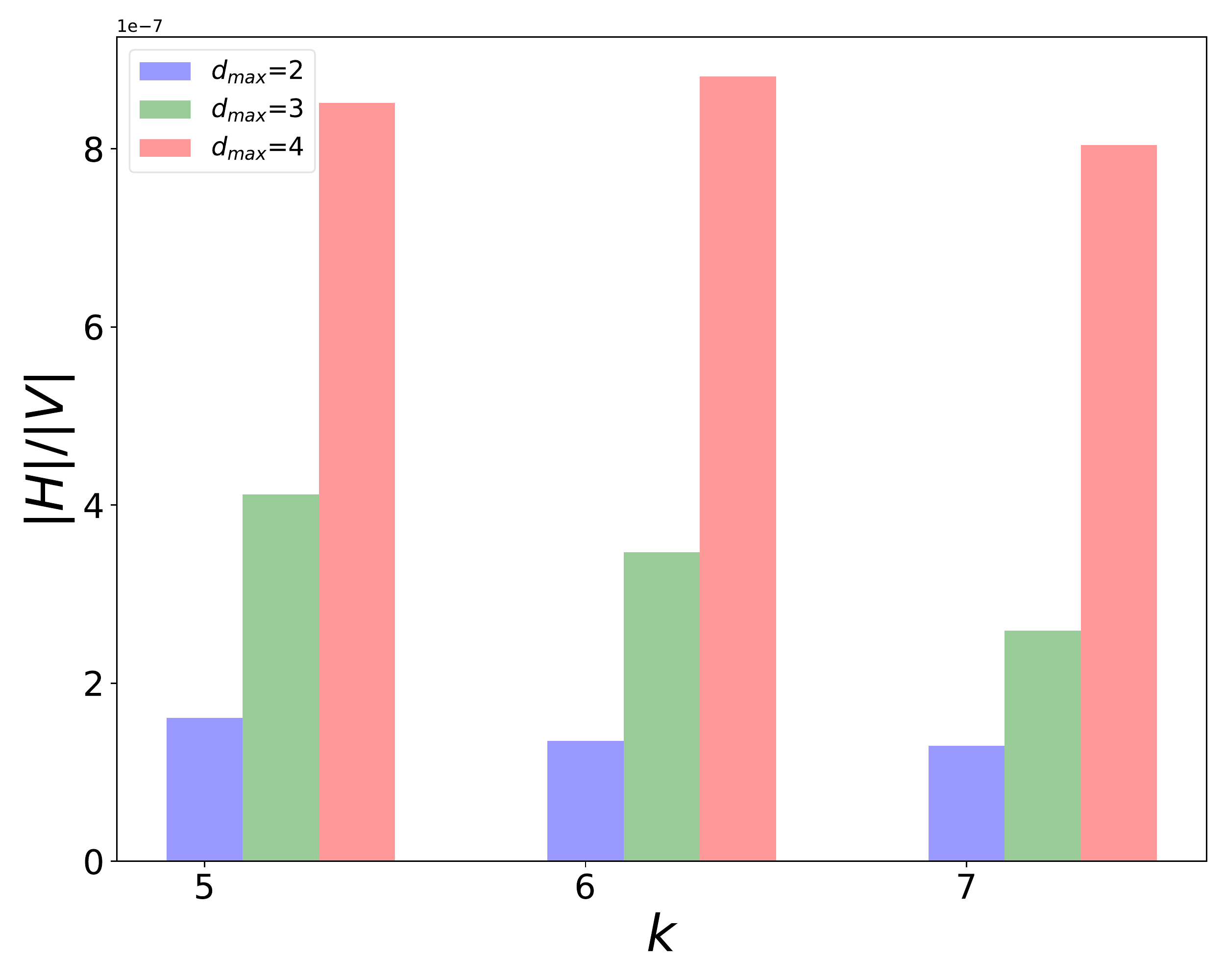}\\	
				\hspace{-4mm}\includegraphics[width=.25\textwidth]{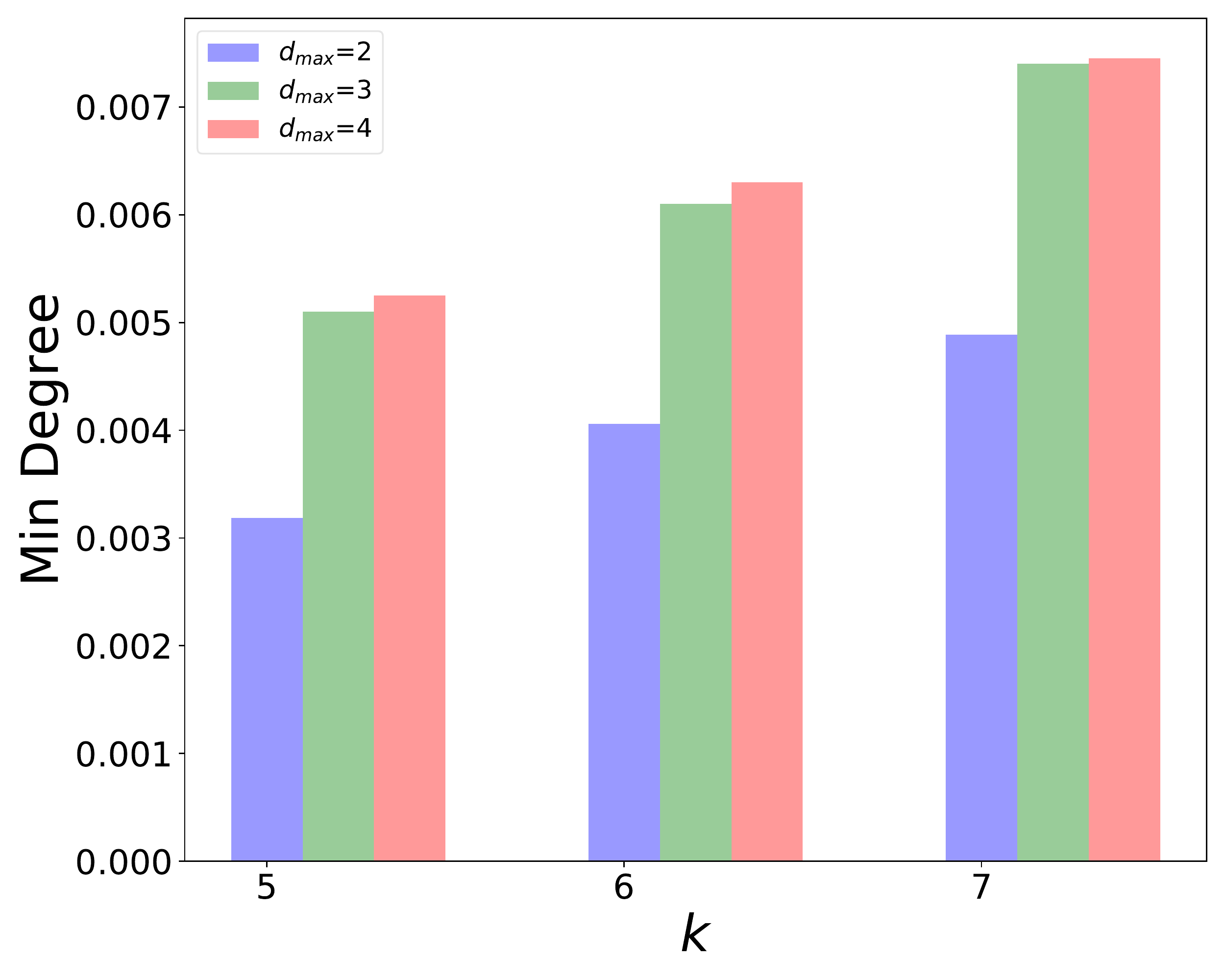}&	\hspace{-4mm}\includegraphics[width=.25\textwidth]{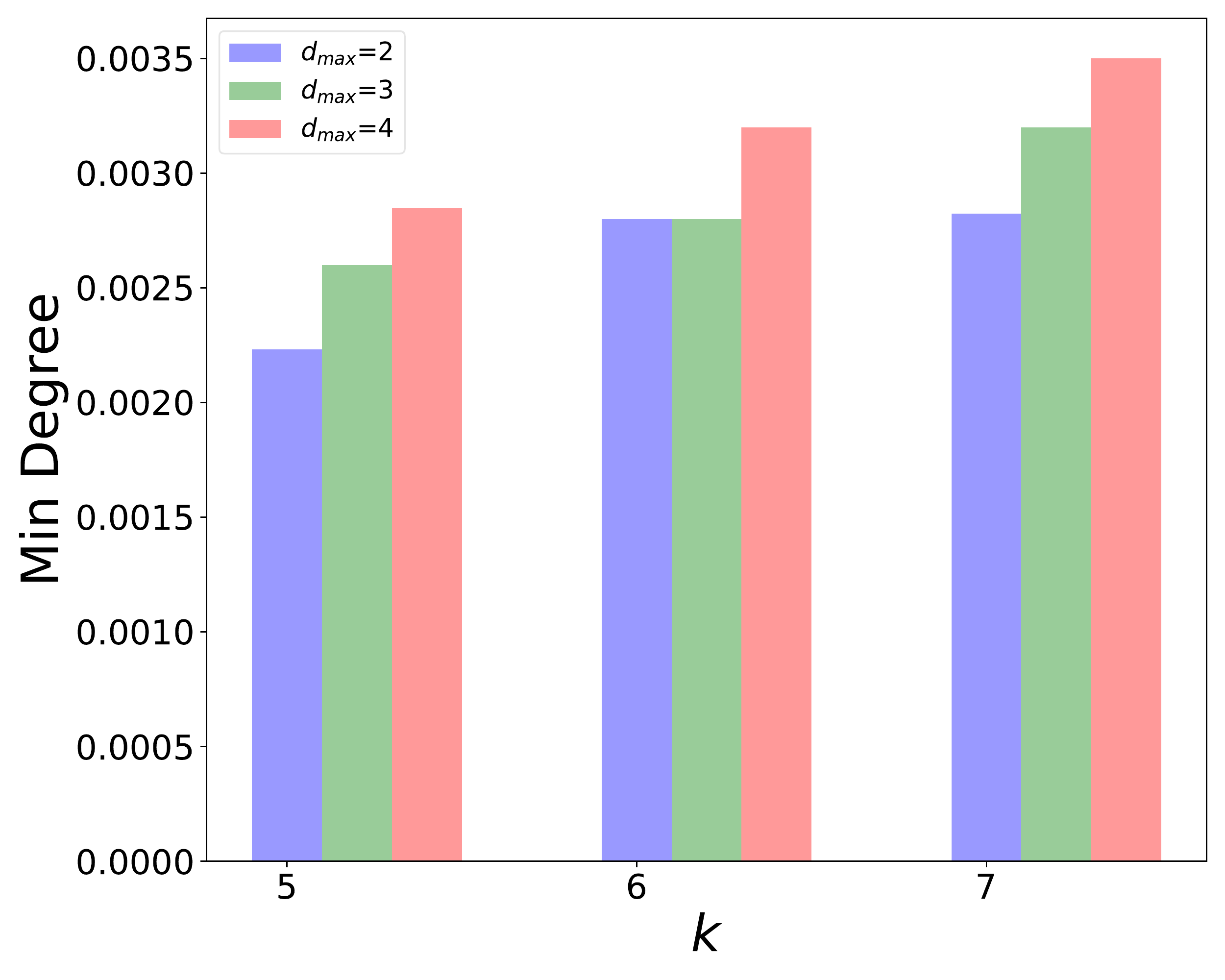}\\	
	\end{tabular}
\vspace{-2mm}
	\caption{Characteristics of the solution on \texttt{youtube} (left) and \texttt{amazon} (right) for \mindeg~with $k \in \{4,6,8\}$ and the maximum diameter threshold $diam_{\max}$ in the range $[2,4]$.}
	\label{fig:dblp-youtube-mindeg}
\vspace{2mm}
\end{figure}

\begin{figure}[t!]
	\begin{tabular}{cc}	
		\hspace{-4mm}\includegraphics[width=.25\textwidth]{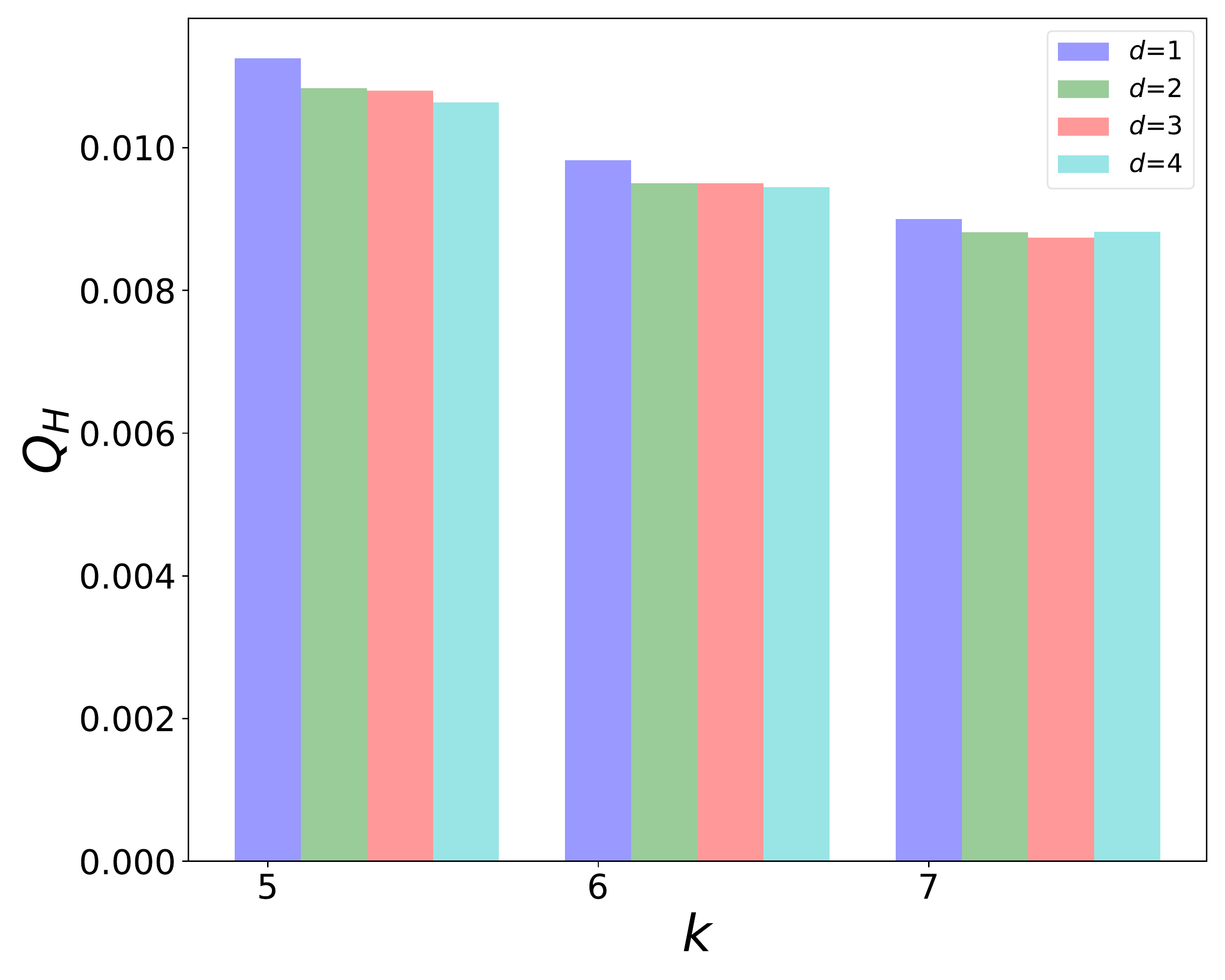}& \hspace{-4mm}\includegraphics[width=.25\textwidth]{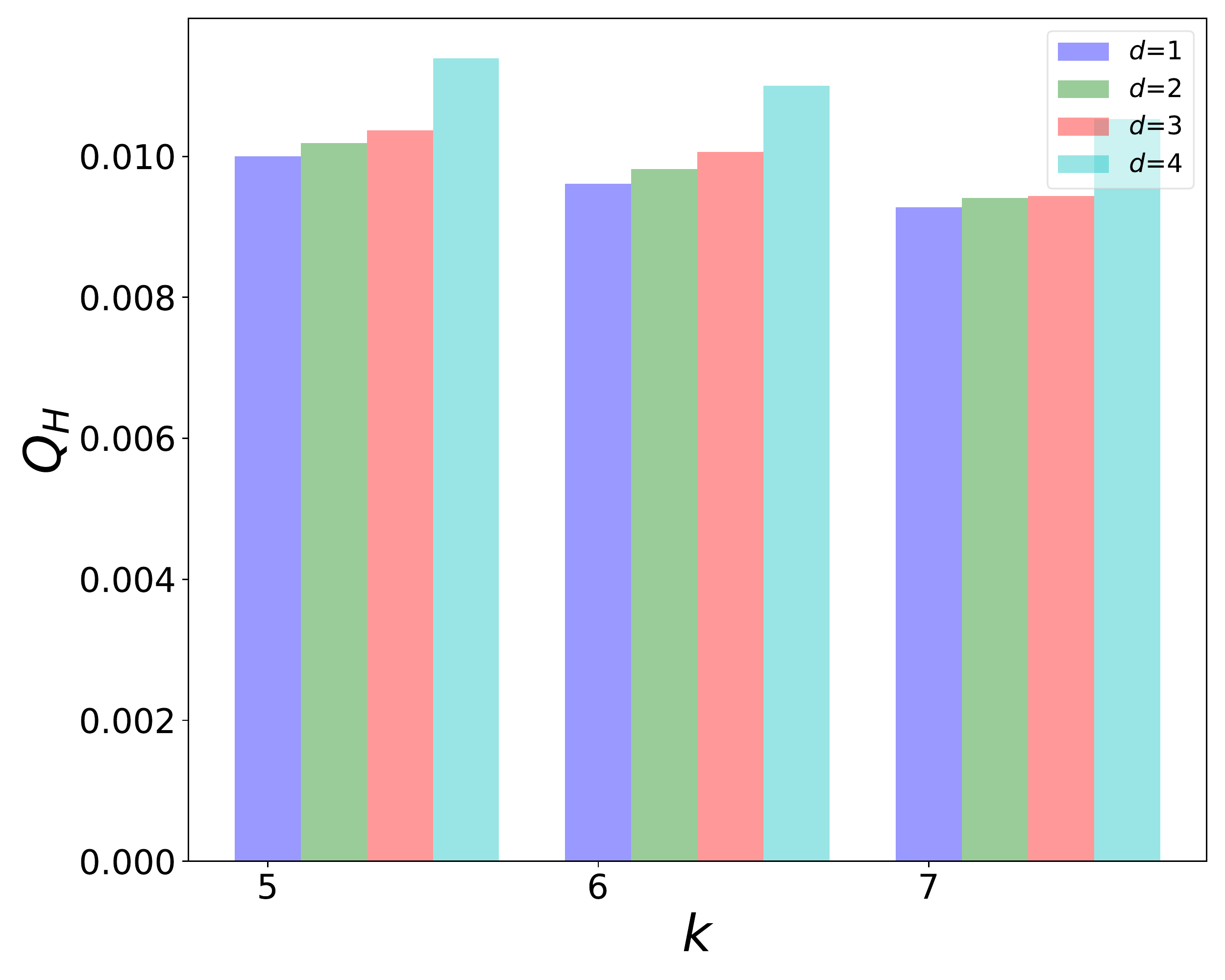}\\
		\hspace{-4mm}\includegraphics[width=.25\textwidth]{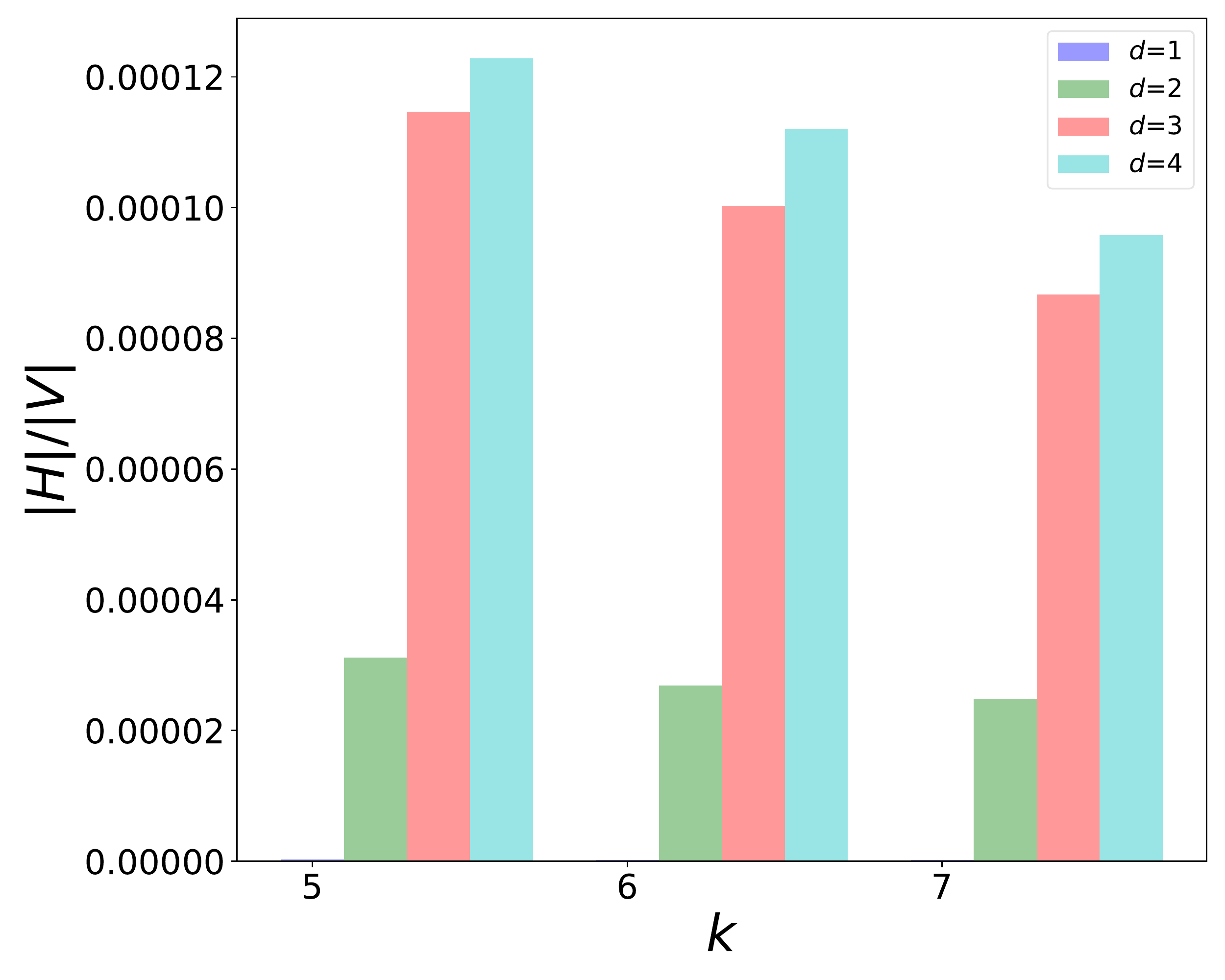}& \hspace{-4mm}\includegraphics[width=.25\textwidth]{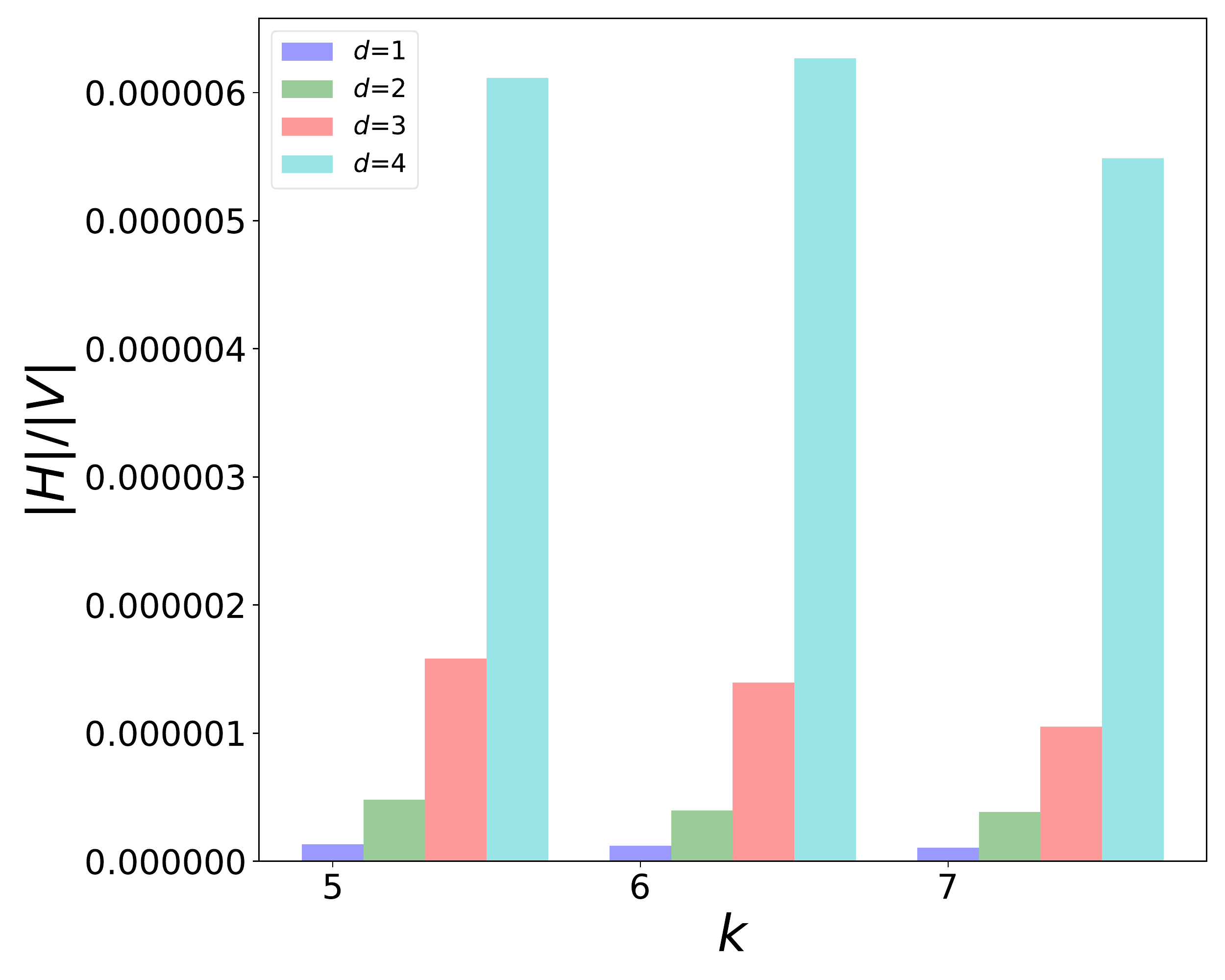}\\
				\hspace{-4mm}\includegraphics[width=.25\textwidth]{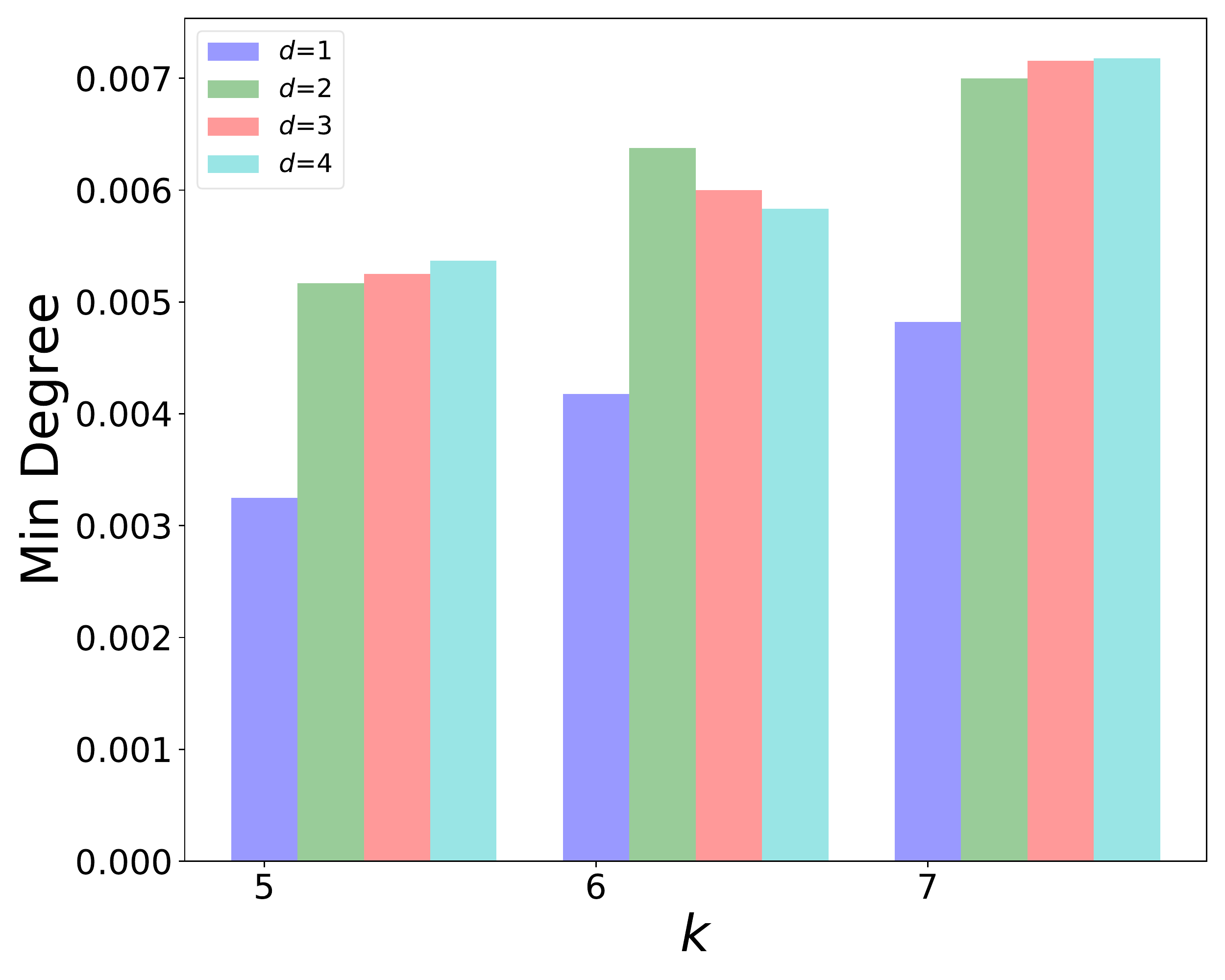}& \hspace{-4mm}\includegraphics[width=.25\textwidth]{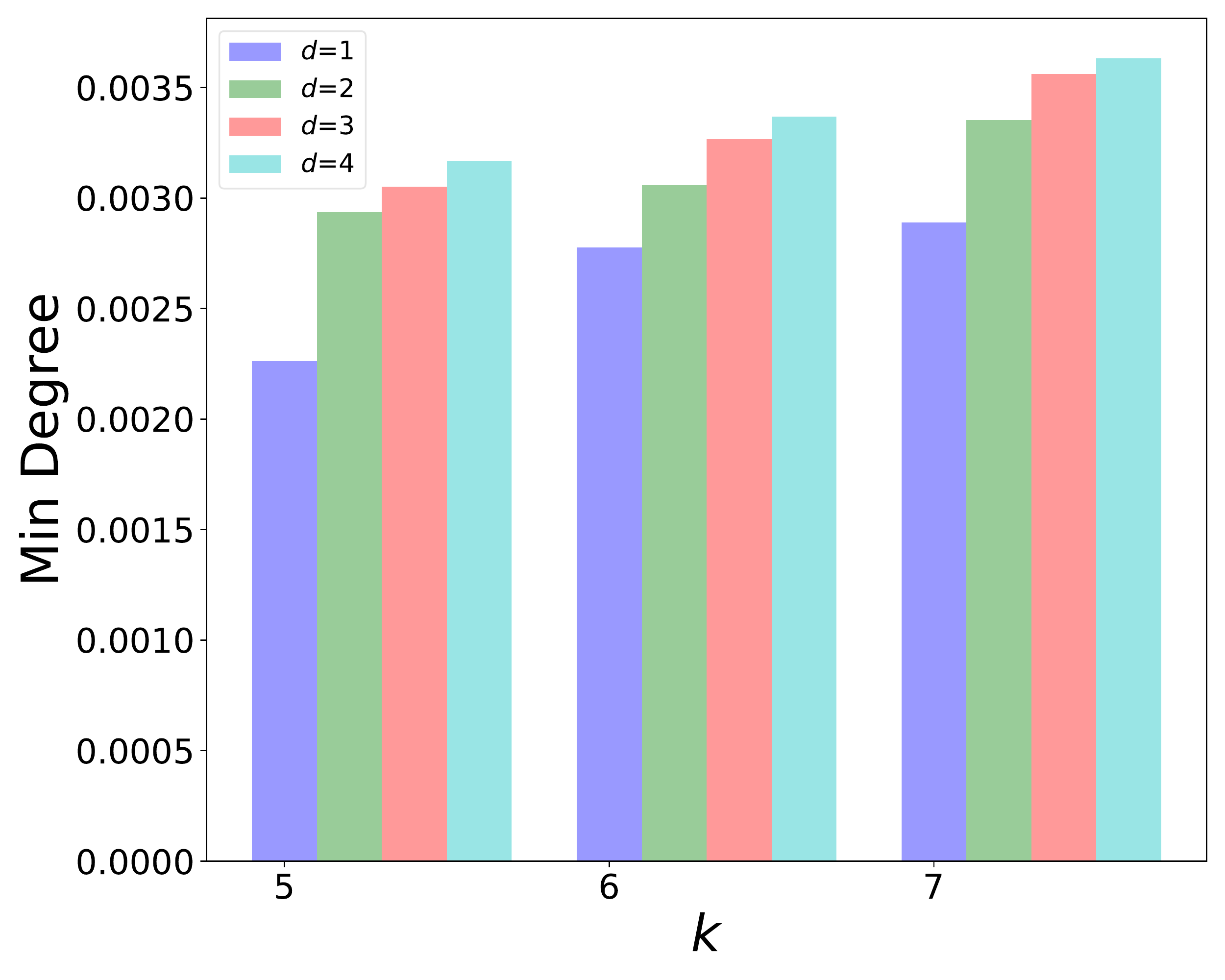}\\
	\end{tabular}
\vspace{-2mm}
	\caption{Characteristics of the solution on \texttt{youtube} (left) and \texttt{amazon} (right) for \mindegdist~with $k \in \{4,6,8\}$ and the threshold on the maximum distance to the query vertices in $[1,4]$.}
	\label{fig:dblp-youtube-mindegdist}
\vspace{2mm}
\end{figure}
The results are reported in Figure \ref{fig:dblp-youtube-mindiam} for \mindiam, Figure \ref{fig:dblp-youtube-mindeg} for \mindeg, and Figure \ref{fig:dblp-youtube-mindegdist} for \mindegdist, over two datasets: \texttt{youtube} and \texttt{dblp}.
For \mindiam~we vary the minimum degree threshold $\delta_{\min}$ in the range $[3,6]$;
for \mindeg~we vary the maximum diameter threshold $diam_{\max}$ in the range $[2,4]$;
for \mindegdist~we vary the threshold on the maximum distance to the query vertices in $[1,4]$.
The first high-level observation is that, for all the three problems studied and under all settings, \emph{increasing the number of allowed outliers produces ``better'' (i.e., more cohesive) solutions}: smaller in size, denser, with higher minimum degree. This was the starting motivation at the basis of this work.
Similar observations can be done w.r.t. the selectivity of constraints. For instance, for \mindiam\ (Figure \ref{fig:dblp-youtube-mindiam}), the higher
is the minimum degree threshold $\delta_{\min}$ the more cohesive the solutions found: i.e., denser and smaller. For \mindeg\ (Figure \ref{fig:dblp-youtube-mindeg}), the smaller is the  maximum diameter threshold $diam_{\max}$, the more cohesive the solutions found. Finally for \mindegdist\ (Figure \ref{fig:dblp-youtube-mindegdist}) imposing a smaller maximum distance to the query vertices, produces more cohesive solutions.

\begin{figure}[t!]
	\begin{tabular}{cc}	
		\hspace{-6mm}\includegraphics[width=.28\textwidth]{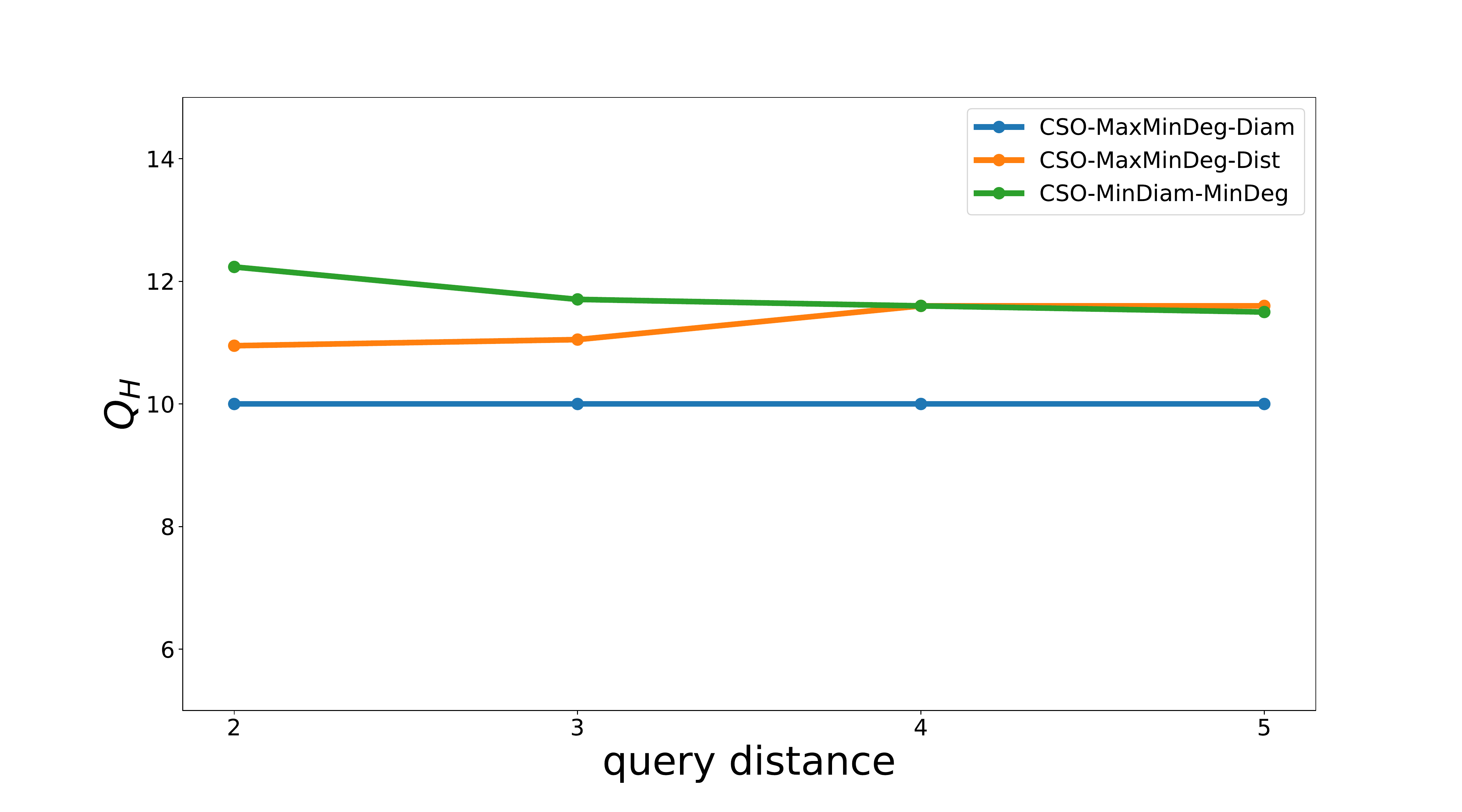}&
		\hspace{-8mm}\includegraphics[width=.28\textwidth]{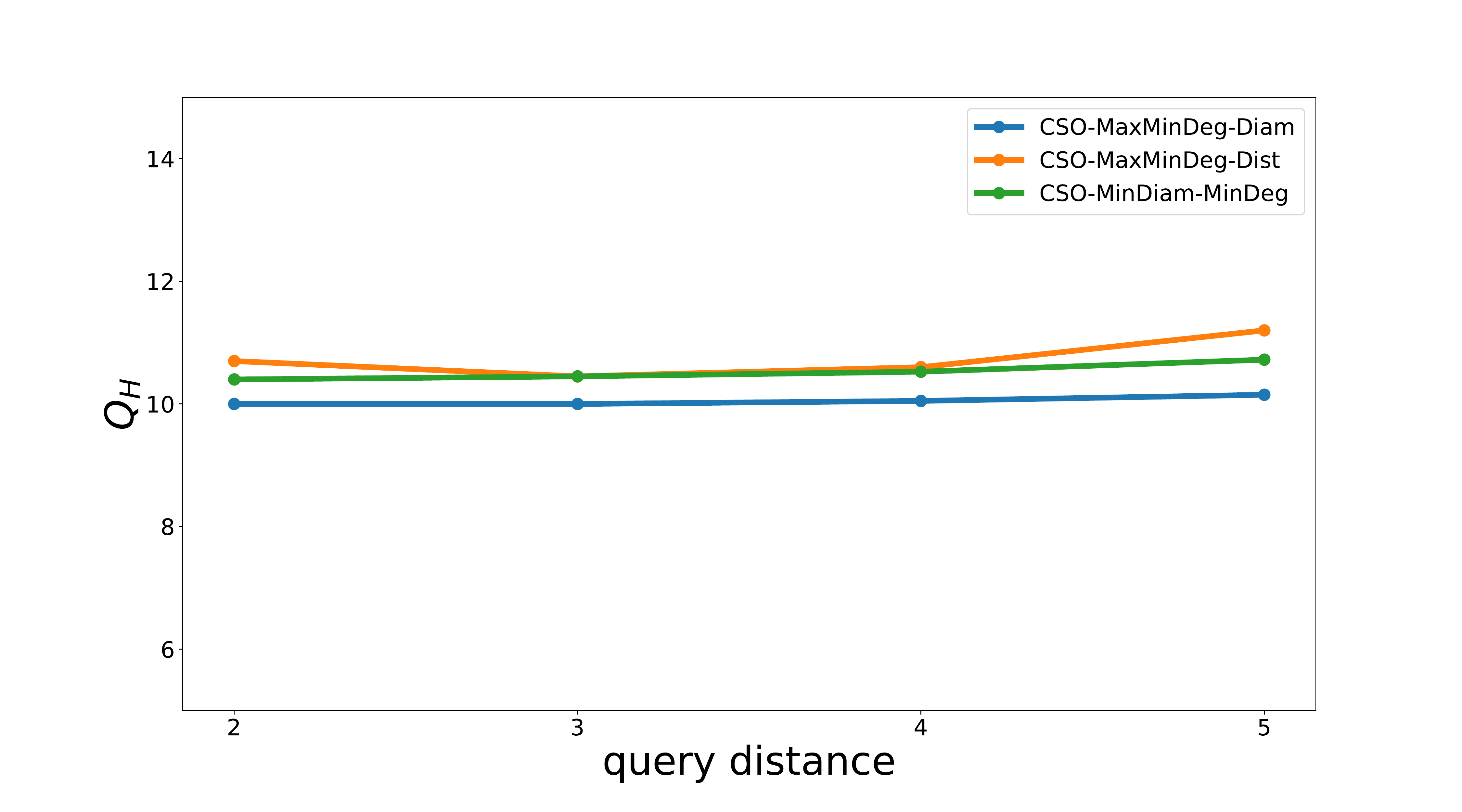}\\
		\end{tabular}
		\vspace{-2mm}
	\caption{ Characteristics of the solution on \texttt{amazon} (left) and \texttt{dblp} (right) graphs with $k=5$ varying the maximum distance among the query vertices inside the same community for \mindiam~with $\delta_{min}=4$, \mindeg~with $diam_{max}=4$, and \mindegdist~with $d_{max}=4$.}
	\label{fig:vd-dblp-ama}
	
\vspace{2mm}
	
	\begin{tabular}{cc}	
	\hspace{-6mm}\includegraphics[width=.28\textwidth]{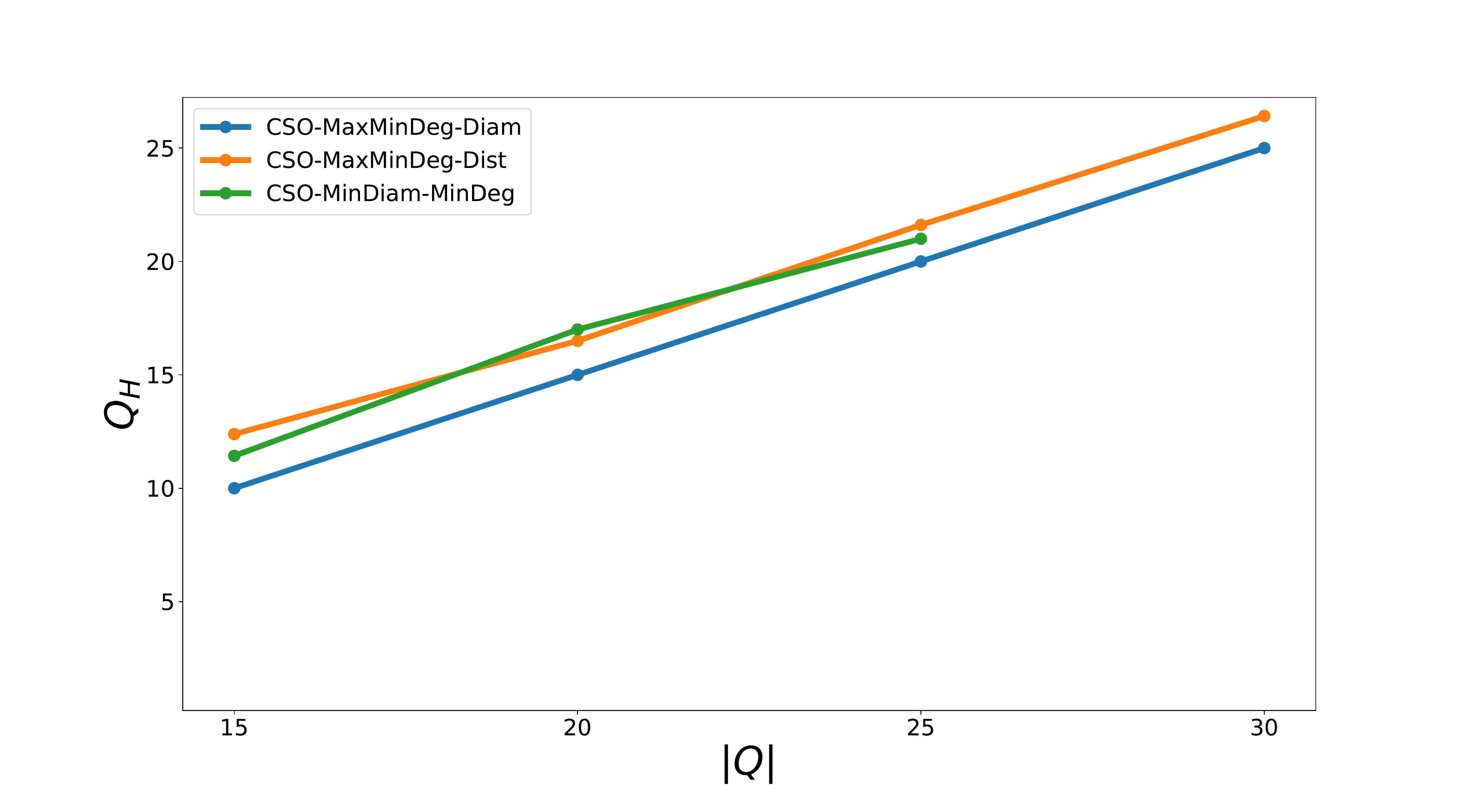}&
	\hspace{-8mm}\includegraphics[width=.28\textwidth]{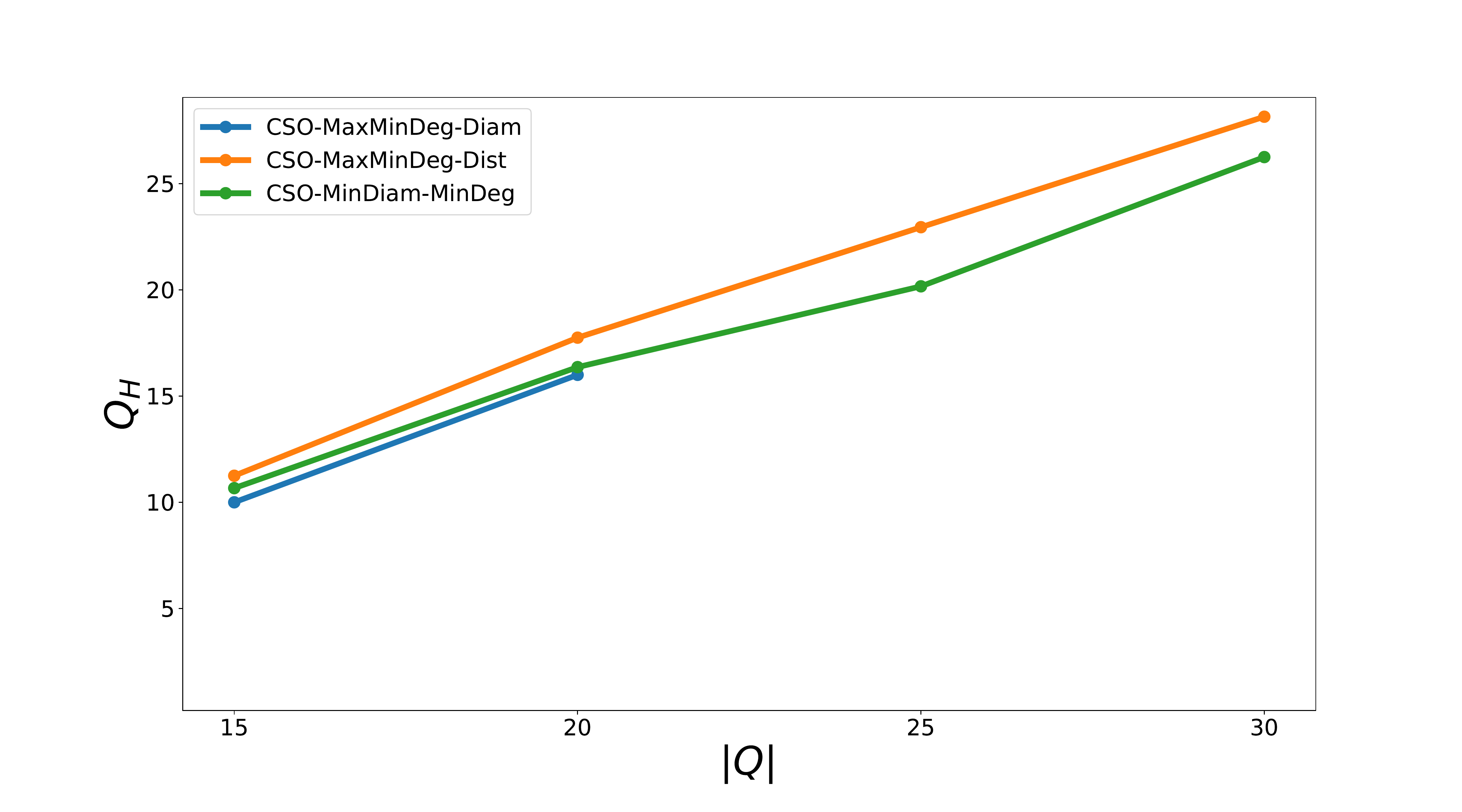}\\
\end{tabular}
	\vspace{-2mm}
	\caption{Characteristics of the solution on \texttt{amazon} (left) and \texttt{dblp} (right) graphs with $k=5$ varying the number of query vertices inside the same community for \mindiam~with $\delta_{min}=4$, \mindeg~with $diam_{max}=4$, and \mindegdist~with $d_{max}=4$. }
	\label{fig:vq-dblp-ama}
\vspace{2mm}
\end{figure}

\begin{figure}[t!]
	\begin{tabular}{cc}	
		\texttt{\mindiam} &	\texttt{\mindeg}   \\ \hspace{-4mm}\includegraphics[width=.25\textwidth]{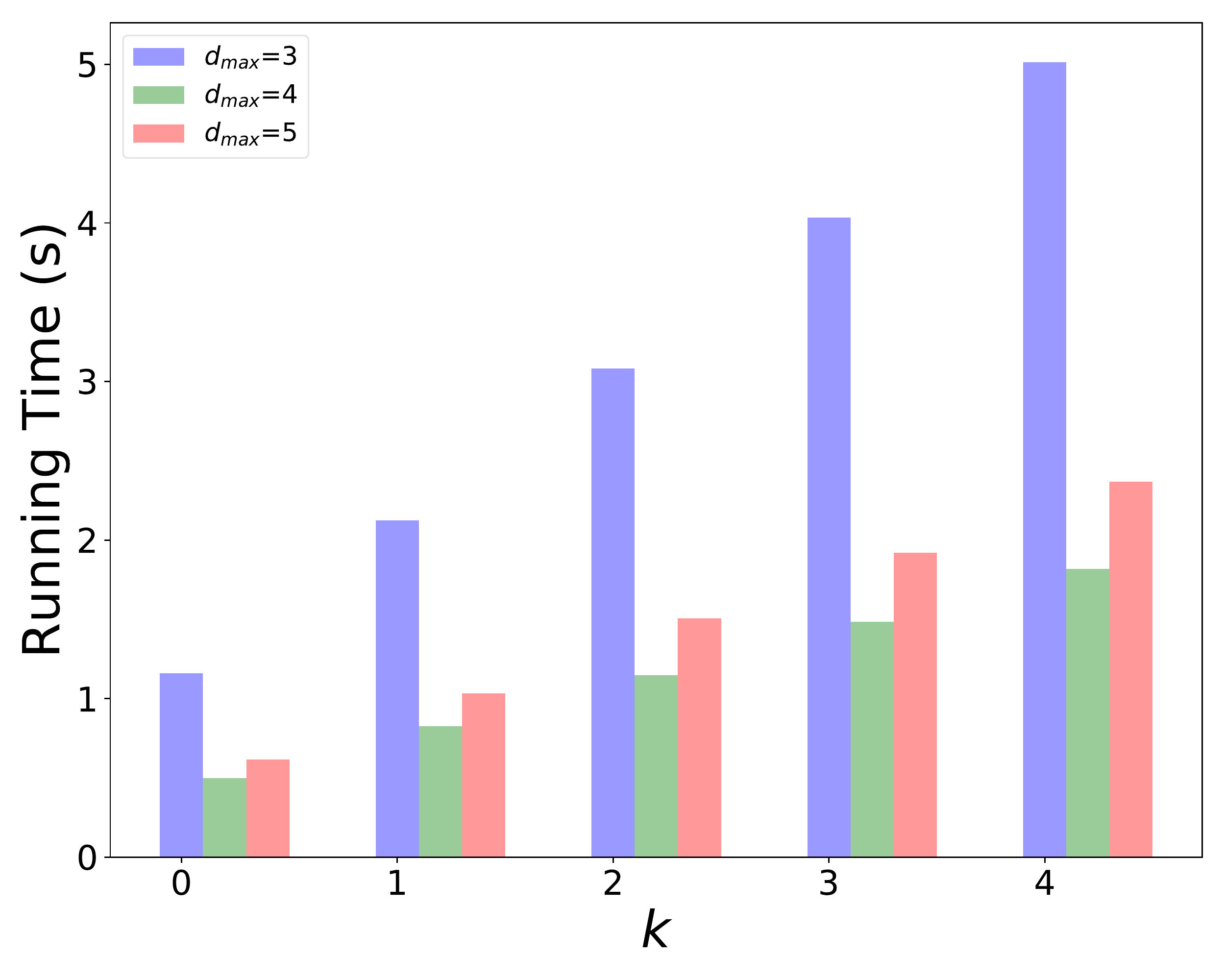}& \hspace{-4mm}\includegraphics[width=.25\textwidth]{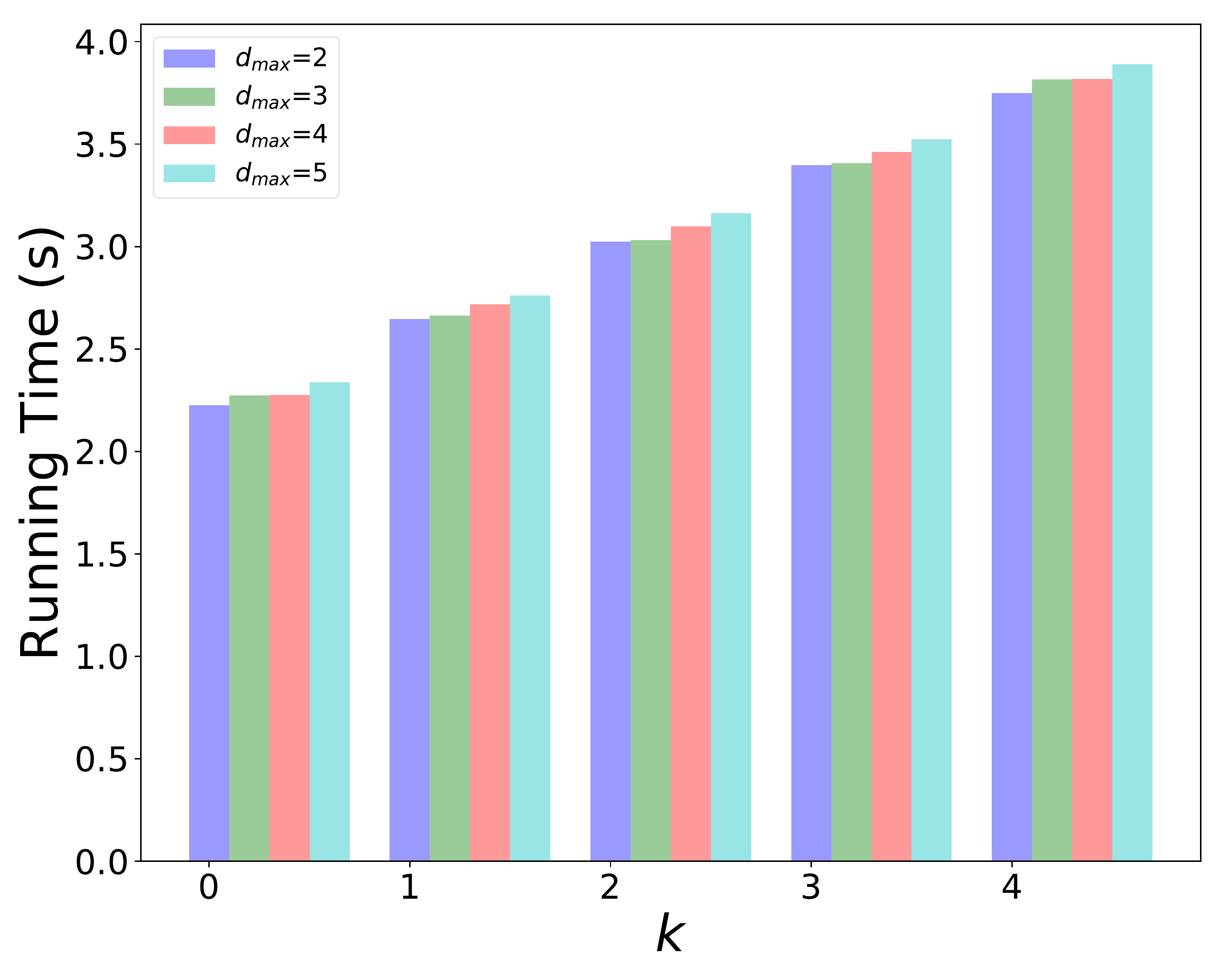}\\
	\end{tabular}
	\vspace{-2mm}
		\caption{Runtime on \texttt{amazon} graph with $n = 5$ and $m = k \in [0,4]$ for different values of the constraints.}
	\label{fig:runtime2}
\vspace{2mm}
\end{figure}

\begin{figure}[t!]
	\begin{tabular}{cc}	
		\multicolumn{2}{c}{\texttt{{\mindegdist}}}   \\ \hspace{-4mm}\includegraphics[width=.25\textwidth]{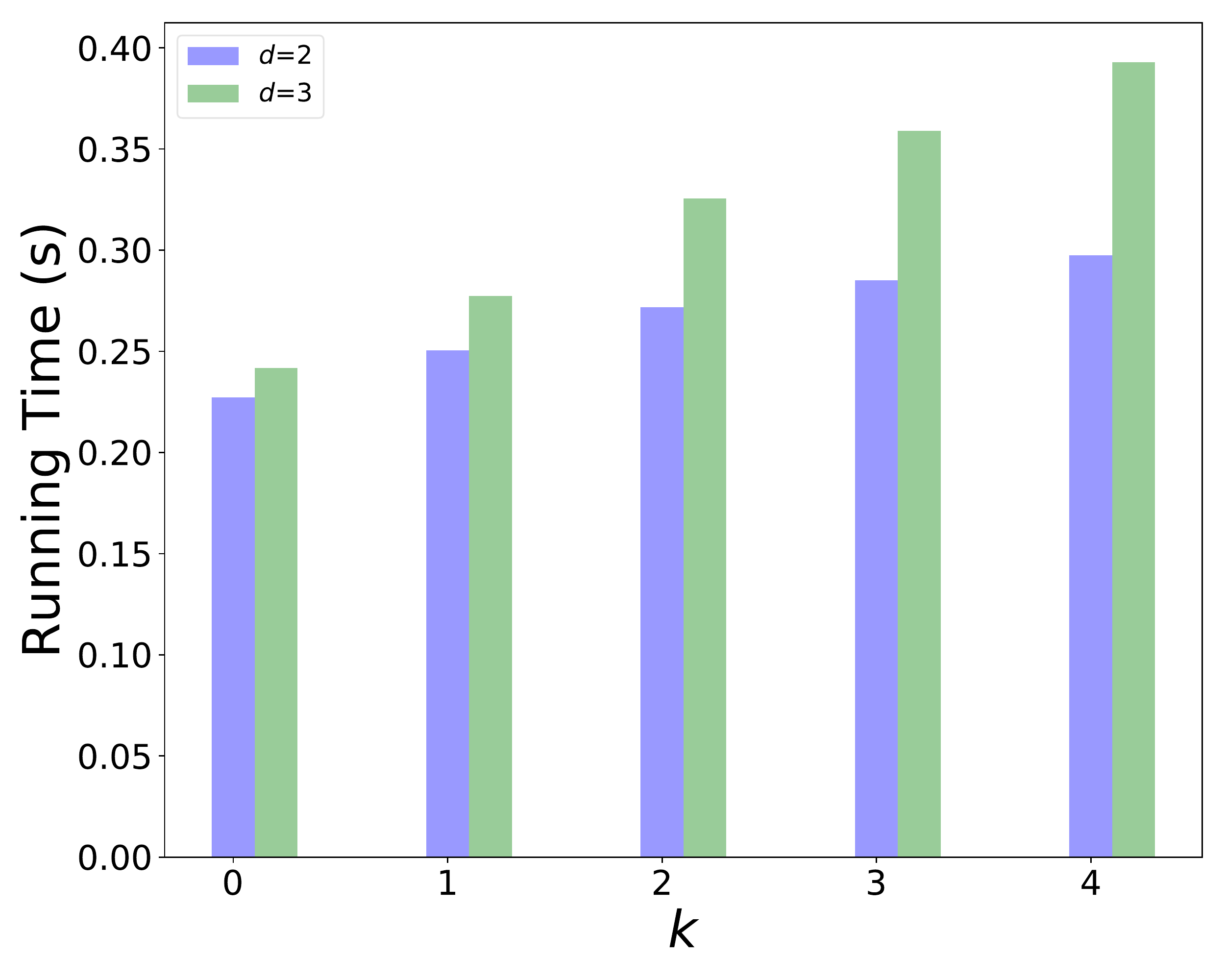}\ & \hspace{-4mm}\includegraphics[width=.25\textwidth]{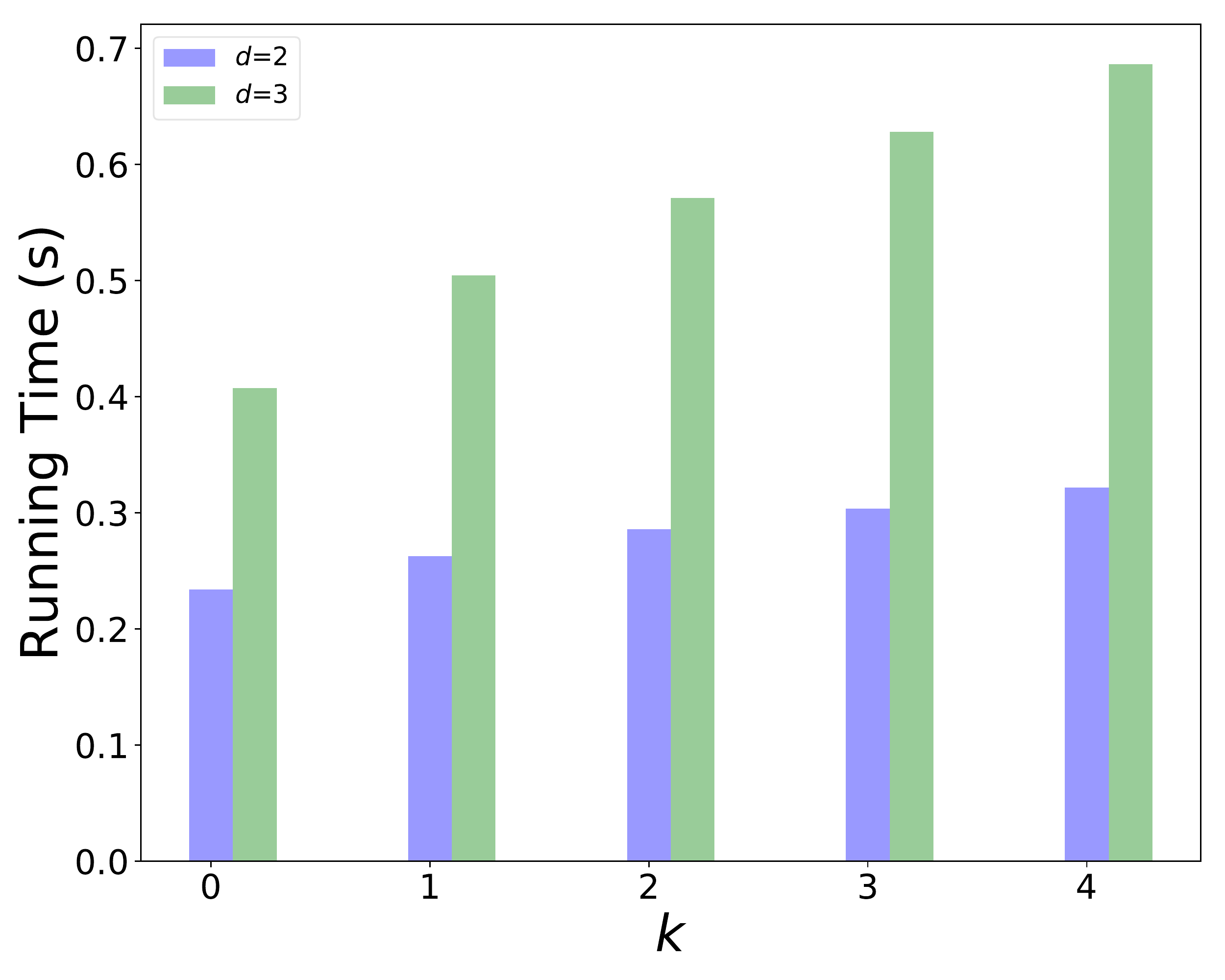}\\
	\end{tabular}
	\vspace{-2mm}
	\caption{Runtime on \texttt{amazon} (left) and \texttt{dblp} (right) graph with $n = 5$ and $m = k \in [0,4]$ for different values of the constraints.}
	\label{fig:runtime}
\vspace{2mm}
\end{figure}


\spara{Characterization with different query sets.}
We next study how the solutions change varying the characteristics of the query set $Q$: in particular the number of query vertices and the distance among query vertices.
In Figure~\ref{fig:vd-dblp-ama} we select $n = 10$ vertices in the same community and $m = 5$ outliers and we varying the distance among the query vertices. We set the input parameter $k =5$.
We notice that the increase the distance has a low effect on the value of $Q_H$.
In Figure~\ref{fig:vq-dblp-ama} we set $m=5$ and we vary the parameter $n$ selecting different sets of query vertices (without any distance constraint): in particular we use $n \in \{10,15,20,25\}$.
On the $x$-axis we report the size of the query set $|Q| = n + m$.
In this case the number of query vertices in the solution grows linearly with respect to the total number of query in input.

\spara{Runtime and parallelization}
Finally, we report runtime performance of the three algorithms in Figure \ref{fig:runtime2} and in Figure~\ref{fig:runtime}.
We can observe that the runtime generally grows with the value of $k$. Moreover, for all the three problems as the constraints are more selective (and thus the solutions more compact), the runtime decreases.
We also tested runtime on the largest dataset, i.e., \texttt{ljournal}, containing approximatively 4M vertices and 34M edges. For all problems we produce the same workload of 20 query sets using $n = 5$ vertices belonging to the same community, $m = 4$ vertices belonging to different communities, and setting the input parameter $k = 4$.
For \mindiam~($\delta_{min}=20$) the runtime was 8333.5 seconds (slightly more than 2 hours), while for \mindeg~($diam_{max}=2$) the runtime was  56.8 seconds, and for \mindegdist~($d_{max}=2$) was 21.63 seconds.
We can improve the scalability of the algorithms for  \mindiam~(respectively, \mindeg) by straightforward parallelization of the subgraph extraction in the following way: we assign to each processor a query vertex, run \amindeg~(respectively \amindiam) in parallel on each processor and compute the best solution among all the processors.

\section{Conclusions}
\label{sec:conclusions}
\vspace{-1.5mm}

We study several variants of the community search problem, where we are allowed to drop up to $k$ query vertices. We adopt three measures of cohesiveness: the minimum degree, the diameter, and the maximum distance with a query vertex. This leads to the formulation of three natural optimization problems, for each of which we either develop an efficient exact algorithm or prove its hardness and develop an approximation algorithm. Our work is the first one to propose algorithms with strong theoretical guarantees for the problem of community search with outliers.
Our experimental evaluation  shows the effectiveness of our approach in identifying outlier query vertices, while showing that much more cohesive solutions can be found when we are allowed to remove outliers.

\spara{Acknowledgments.}
Francesco Bonchi acknowledges support from Intesa Sanpaolo Innovation Center.
Lorenzo Severini is with UniCredit Group.
Mauro Sozio has been working in the frame of a cooperation between Huawei Technologies France SASU and Telecom Paris (Grant no. YBN2018125164).
The content of this paper and the views expressed here, are solely responsibility of the authors. 
The funders had no role in study design, preparation of the manuscript or decision to publish.

\bibliographystyle{abbrv}
\bibliography{references,refs_short,biblio_short}

\end{document}